\documentclass[aps,superscriptaddress,showpacs,twocolumn,amsfonts,amsmath]{revtex4}

\usepackage[final]{graphicx}
\usepackage{amsmath}
\usepackage{latexsym}
\usepackage{float}
\usepackage{amssymb}

\newlength\figurewidth
\AtBeginDocument{\setlength\figurewidth{.9\linewidth}}

\usepackage{color}
\let\omp\marginpar\relax\def\marginpar#1{\omp{\color{red}#1}}

\newcommand{\eq}[1]{Eq.~(\ref{#1})}
\newcommand{\fig}[1]{Fig.~\ref{#1}}
\newcommand{\olcite}[1]{Ref.~[\onlinecite{#1}]}

\newcommand{\kb}{k_{\rm B}T}
\newcommand{\etacv}{\etac^{\rm v}}
\newcommand{\etacl}{\etac^{\rm l}}

\newcommand{\etac}{\eta_{\rm c}}
\newcommand{\etaccr}{\eta_{\rm c,cr}}
\newcommand{\etapr}{\eta_{\rm p}^{\rm r}}
\newcommand{\etaprcr}{\eta_{\rm p,cr}^{\rm r}}

\begin{document}

\title{Phase diagram and structure of colloid-polymer mixtures confined between
walls}

\date{\today}

\author{R. L. C. Vink$^{1,2}$, A. De Virgiliis$^1$, J.
Horbach$^1$, and K. Binder$^1$ \\
$^1$ Institut f\"ur Physik, Johannes--Gutenberg--Universit\"at
Mainz, Staudinger Weg 7, D-55099 Mainz, Germany \\
$^2$ Institut f\"ur Theoretische Physik II, 
Heinrich--Heine--Universit\"at D\"usseldorf, Universit\"atsstra{\ss}e 1,
D--40225 D\"usseldorf, Germany}

\begin{abstract}

The influence of confinement, due to flat parallel structureless walls, on 
phase separation in colloid-polymer mixtures, is investigated by means of 
grand-canonical Monte Carlo simulations. Ultra--thin films, with 
thicknesses between $D=3-10$ colloid diameters, are studied. The 
Asakura-Oosawa model [J. Chem. Phys. \textbf{22}, 1255 (1954)] is used to 
describe the particle interactions. To simulate efficiently, a ``cluster 
move'' [J. Chem. Phys. \textbf{121}, 3253 (2004)] is used in conjunction 
with successive umbrella sampling [J. Chem. Phys.~\textbf{120}, 10925 
(2004)]. These techniques, when combined with finite size scaling, enable 
an accurate determination of the unmixing binodal. Our results show that 
the critical behavior of the confined mixture is described by 
``effective'' critical exponents, which gradually develop from values near 
those of the two-dimensional Ising model, to those of the 
three-dimensional Ising model, as $D$ increases. The scaling predictions 
of Fisher and Nakanishi [J. Chem. Phys.~\textbf{75}, 5875 (1981)] for the 
shift of the critical point are compatible with our simulation results. 
Surprisingly, however, the colloid packing fraction at criticality 
approaches its bulk ($D \to \infty$) value non-monotonically, as $D$ is 
increased. Far from the critical point, our results are compatible with 
the simple Kelvin equation, implying a shift of order $1/D$ in the 
coexistence colloid chemical potential. We also present density profiles 
and pair distribution functions for a number of state points on the 
binodal, and the influence of the colloid-wall interaction is studied.

\end{abstract} 


\maketitle

\section{Introduction}

Confinement of fluids in nanoscopic capillaries is a problem that has 
received longstanding attention. An interesting interplay occurs between 
surface effects at the confining walls, such as wetting or drying 
\cite{1,2,3,4}, and finite size effects due to the finite capillary width. 
This interplay leads to a host of intriguing phenomena, such as capillary 
condensation or evaporation \cite{5,6,7,8,9}, and a crossover in critical 
behavior from three-dimensional (3d) to two-dimensional (2d) Ising 
character \cite{4,5,8,10,11}. Apart from fundamental theoretical importance, 
understanding the structure and phase behavior of confined fluids is also 
a prerequisite for targeted applications in microfluidic and nanofluidic 
devices, which are becoming increasingly more relevant \cite{7,12,13,14}.

However, for fluids consisting of small molecules, complications such as the
atomistic corrugation of the confining walls \cite{15}, impurity atoms at the
walls, roughness of the walls due to surface steps, dislocations in the crystal
structure, and so forth, may have a profound effect on the phenomena mentioned
above. This is especially true when the width of the slit pore is of the order
of a few nanometers, in which case a quantitative understanding becomes
difficult. In this respect, colloidal dispersions, containing colloidal
particles with diameters in the micrometer range, possess certain advantages. It
then becomes possible to prepare slit pores which are only a few particle
diameters wide, and yet have walls that are essentially flat on the size scale
of the particles. In addition, interactions between colloidal particles can be
well tuned \cite{16,17,18}. Particularly promising systems of this kind are
colloid--polymer mixtures, since both the bulk phase behavior, and the
interfaces separating the colloid--rich and polymer--rich phases, can be studied
experimentally in detail \cite{19,20,21,22}. Moreover, the Asakura--Oosawa (AO)
model \cite{23,24} provides a simple theoretical framework, which seems to
capture the essential features of such phase--separating systems, and is
moreover well suited for computer simulations \cite{8,25,26,27,28,29,30}.

The aim of this work is to provide precise predictions for the phase diagram and
the structure of the AO model confined between structureless hard walls. We
consider film thicknesses ranging from $D=3$ to $D=10$ colloid diameters. Our
work complements earlier work \cite{27,28}, based on Gibbs ensemble Monte Carlo,
which focused on the non-critical regime of the phase diagram. In this work,
also the critical regime is considered, in order to compare to theoretical
predictions for the shift in critical parameters as function of the film
thickness \cite{5,11}. We consider a colloid-to-polymer size ratio $q =
\sigma_{\rm p}/\sigma_{\rm c} = 0.8$, with $\sigma_{\rm c}$ the colloid diameter
and $\sigma_{\rm p}$ the polymer diameter of gyration, since for this particular
size ratio, accurate information on the bulk and interfacial properties is
available \cite{29,30}. We also explore the effect of an additional weak
repulsion between the walls and the colloids (in addition to the hard
interaction already present). Understanding the combined effect of film
thickness and wall-colloid interaction, is crucial in providing guidance for the
interpretation of possible experiments. At this point, we are aware of only one
experiment on capillary condensation of a colloid--polymer mixture (in a wedge
formed by glass plates \cite{22}), but we hope that the present study will
encourage further experiments. In related previous work carried out by us, in
which a film thickness $D=5$ and purely hard walls were considered, the
crossover from 3d to 2d Ising critical behavior was already discussed \cite{8}.

The outline of this paper is as follows. In Sec.~II, we summarize the most
important theoretical predictions relevant for the interpretation of our
results. In Sec.~III, the AO model is introduced, and details on the simulation
technique are provided. In Sec.~IV, we investigate the phase behavior and the
structural properties of the AO model as the film thickness $D$ is varied, while
Sec.~V considers the influence of the colloid--wall interaction. Finally, in
Sec.~VI, we summarize our main conclusions.

\section{Theoretical background}

We consider a system with an one-component order parameter near its critical
point. Examples of such systems are fluids near their vapor-liquid critical
point, magnetic systems (such as the simple Ising model), and binary mixtures
(such as colloid-polymer mixtures) near their critical point of unmixing. As is
well known, all these systems in 3d bulk belong to the 3d~Ising universality
class \cite{31}. In particular, critical exponents such as the exponent $\nu$
(which characterizes the growth of the order parameter correlation length $\xi$
near the critical point), or the order parameter exponent $\beta$ (which
characterizes the shape of the binodal between the unmixed phases in a
binary mixture) have nontrivial values $\nu \approx 0.630$, $\beta \approx
0.326$ \cite{32,33,34}, rather than the classical mean--field values $\nu_{\rm
MF}= 1/2$, $\beta_{\rm MF} = 1/2$.

When these systems are confined between two identical flat planar structureless
walls a distance $D$ apart, two distinct physical phenomena affecting the phase
transition are expected to occur:
\begin{enumerate} 
\item[(i)] 
The growth of the critical correlations in the direction perpendicular to the
walls is limited by the finite thickness of the film, while critical
correlations can still grow further and unlimited in the two directions parallel
to the film. Therefore, a crossover from 3d~critical behavior toward 2d~critical
behavior (for which the exponents mentioned above take the values $\nu_2=1$ and
$\beta_2=1/8$, assuming 2d~Ising universality \cite{31,35}) is expected. This
crossover is predicted to occur when the temperature-like variable has
approached a relative distance $t$ from the critical point in the bulk such that
$\xi \propto t^{-\nu}$ is of the same order as $D$, implying a crossover
distance of order $t_{\textrm{cross}} \propto D^{-1/\nu}$. This dimensional
crossover in critical behavior also implies an additional enhancement of
fluctuations (in lower dimensionality stronger fluctuations occur). This leads
to a depression of the critical point. Consequently, one expects a shift of the
critical point of the same order, $t_{\textrm{shift}} \propto D^{-1/\nu}$.
\item[(ii)] 
If there is no special symmetry between the coexisting phases, one must expect
that there is an (energetic and/or entropic) preference of the walls for one of
the coexisting phases in comparison to the other. This phenomenon is also well
known to occur away from the critical point. For example, hydrophilic walls of a
thin slit capillary are known to lead to ``capillary condensation'' of an
undersaturated vapor in the capillary \cite{36}. Of course, for hydrophobic
rather than hydrophilic walls, also the opposite effect (``capillary
evaporation'') may occur.
\end{enumerate}

Clearly, a quantitative prediction of the magnitude, and sometimes even the
sign, of these effects, requires detailed knowledge of the forces between the
walls and the particles confined by them. However, since phase equilibrium is
always shifted due to surface corrections to the relevant thermodynamic
potentials, we expect a shift in the coexistence chemical potential in the thin
film, relative to its value in the bulk, of order $1/D$. This is the so-called
``Kelvin equation'' \cite{4,5,6,7}, which is expected to be valid away from the
critical point. In contrast, the corresponding shift in the chemical potential
at criticality is more subtle. This becomes most transparent for the case of an
Ising magnet (or, equivalently, the lattice gas model), where in the bulk there
is a symmetry between the coexisting ``spin up'' and ``spin down'' phases.
However, in the thin film, this symmetry may be broken by a surface magnetic
field $H_1$ \cite{4,5,11,37,38}. While phase coexistence in the bulk occurs for
bulk magnetic field $H=0$, coexistence in the thin film now requires a nonzero
bulk field (with a sign opposite to that of the surface field $H_1$). 

It has been shown that in the presence of the surface field, the singular 
part of the free energy per spin in this Ising model should scale as 
\cite{4,5,11,37,38}
\begin{eqnarray}\label{eq1}
  f_{\textrm{sing}}(D,T,H,H_1) = \hspace{4cm} \\ |t|^{2-\alpha}  
  \tilde{f}_\pm(D|t|^\nu, H |t|^{-\Delta}, H_1|t|^{-\Delta _1}), \nonumber
\end{eqnarray}
with $\alpha \approx 0.110$ \cite{32,33,34} the critical exponent of the 
specific heat, $\Delta \approx 1.56$ \cite{32,33,34} the so-called ``gap 
exponent'' which characterizes the bulk equation of state, and $\Delta_1 
\approx 0.47$ \cite{38,39,40,41} its surface analogue. The scaling 
functions $\tilde{f}_\pm$, where the signs refer to the sign of $t$, are 
discussed in more detail in Refs.~\cite{5,11}. Here, we only infer from 
Eq.~(\ref{eq1}) that the shift in the critical temperature should read 
as \cite{5}
\begin{eqnarray}\label{eq2} 
  t_{\rm shift} &\equiv& \Delta T_{\rm c} / T_{\rm c}(\infty) \\
  &=& \left[ T_{\rm c}(D,H_1)-T_{\rm c}(\infty) \right]/T_{\rm c}(\infty) \nonumber \\
  &=& -D^{-1/\nu} X_{\rm c}(H_1D^{\Delta _1/\nu}), \nonumber 
\end{eqnarray}
and similarly for the shift in the bulk magnetic field required to 
establish coexistence again
\begin{eqnarray}\label{eq3}
  \Delta H_{\rm c} &\equiv& H_{\rm c}(D,H_1) \\
  &=& -D^{-\Delta/\nu}Y_{\rm c}(H_1D^{\Delta_1/\nu}) \nonumber \\ 
  &\stackrel{H_1 \rightarrow 0}{\propto}& - H_1 D^{(\Delta_1-\Delta)/\nu}. \nonumber
\end{eqnarray}
Here, $X_{\rm c}$ and $Y_{\rm c}$ are again scaling functions, and $T_{\rm 
c}(\infty)$ denotes the critical temperature in the bulk ($D \to \infty$) 
system. Note that for small $H_1$, we expect that $X_{\rm c}$ tends to a 
constant, while $Y_{\rm c}$ becomes a linear function of its argument. This 
limit is relevant in the present case, since we deal with rather small 
film thicknesses $D$.

Considering now mixtures of colloids (c) and polymers (p) in the
grand--canonical ensemble, chemical potentials, $\mu_{\rm c}$ and
$\mu_{\rm p}$, of colloids and polymers, respectively, as well as
temperature $T$ and system volume $V$ become the relevant independent
thermodynamic variables. In the framework of the AO model \cite{23,24},
the dependence on temperature only enters via the fugacities $z_{\rm
c}=\exp(\mu_{\rm c}/\kb)$, and $z_{\rm p}=\exp(\mu_{\rm p}/\kb)$. There
is no other explicit temperature dependence in this model. Considering
the fugacity of the polymers as a temperature--like variable, bulk phase
coexistence between a phase rich in colloids (analogous to a liquid in
the liquid-vapor transition) with colloid packing fraction $\etacl$, and
a phase poor in colloids (analogous to the vapor) with colloid packing
fraction $\etacv < \etacl$, occurs on the line $z^{\rm coex}_{\rm c}
(z_{\rm p})$. The latter is determined by the equation $\mu_{\rm c}/\kb =
\mu_{\rm c}^{\rm coex}(z_{\rm p})/\kb$ in the plane of variables ($z_{\rm
p}, z_{\rm c}$). The variable $\Delta\mu = \mu_{\rm c} - \mu_{\rm
c}^{\rm coex}$ thus plays a role analogous to that of the magnetic
field $H$ in the Ising model, and $\etacl(z_{\rm p})-\etacv(z_{\rm p})$
corresponds to $2m_{\rm sp}(T)$ for $H\rightarrow 0^+$, with $m_{\rm
sp}(T)= - (\partial f_{\textrm{sing}} / \partial H)_T$ the spontaneous
magnetization or order-parameter.

While in the Ising model there is a symmetry with respect to the sign of 
$H$, no corresponding symmetry with respect to the sign of $\Delta \mu$ 
occurs in colloid--polymer mixtures, of course. In the Ising model, the 
two branches $\pm m_{\rm sp}(T)$ of the binodal, corresponding to positive and 
negative fields, are equivalent; this is the particle--hole symmetry of 
the lattice gas. Obviously, no such symmetry exists in {\it asymmetric} 
mixtures. Therefore, the above identification of variables holds only to a 
first approximation, and for a more precise discussion of critical 
phenomena one needs to consider ``field mixing'' effects \cite{42}. The 
temperature--like variable $t$ should then be taken parallel to the curve 
$z_{\rm c}^{\rm coex}(z_{\rm p})$ at the point $z_{\rm c}^{\rm 
crit}(z_{\rm p} = z_{\rm p}^{\rm crit})$, and the field--like variable $H$ 
becomes a suitable linear combination of the variables $\Delta \mu$ and 
$z_{\rm p} - z_{\rm p}^{\rm crit}$ \cite{29,30,42}.

Even more intricate becomes the identification of the variable 
corresponding to $H_1$, the local magnetic field coupling to the spins at 
the surface of the Ising magnet. The case $H_1=0$ physically means 
``neutral walls'', in which case the walls prefer neither the 
``spin-up''-phase nor the ``spin-down''-phase of the ferromagnet. It is 
not at all obvious which choice of surface interactions would correspond 
to such a ``neutral wall'', preferring neither the colloid-rich phase, nor 
the polymer-rich phase of our model. Of course, a completely analogous 
difficulty occurs in studies of capillary condensation and/or surface 
critical phenomena associated with the vapor--liquid critical point of 
ordinary fluids \cite{43,44,45}. We conclude that, in general, whatever 
the choice for the strength and range of the particle--wall interaction, 
it will likely correspond to a situation $H_1 \neq 0$, but we have no 
possibility to predict the strength (and even the sign) of $H_1$ 
beforehand. In principle, a careful analysis of order parameter profiles 
at the bulk critical point of a very ``thick'' film (which approximates a 
semi--infinite system with two surfaces) for various wall--particle 
interactions could provide insight into how to realize a situation with 
$H_1=0$ at criticality \cite{38}. This, however, gives no guarantee that, 
for the same choice of interactions, one also has $H_1=0$ outside of the 
critical region.

On the other hand, if our simulations display only very small shifts of 
$\mu_{\rm c}^{\rm coex}(z_{\rm p}) / \kb$ with decreasing film thickness, 
one may assume that $H_1$ in our model is indeed small. In this case, 
Eqs.~(\ref{eq2}) and (\ref{eq3}) still hold. However, if the data indicate 
that this is not the case, and rather the inverse limit $H_1 
D^{\Delta_1/\nu} \gg 1$ is reached instead, simple power laws would again 
result
\begin{equation}\label{eq4}
  t_{\rm shift} = -D^{-1/\nu} X_{\rm c}(\infty), \; 
  \Delta H_{\rm c} = -D^{-\Delta/\nu} Y_{\rm c}(\infty).
\end{equation}
While the first power law of Eq.~(\ref{eq4}) is the same as before (only 
the constant of proportionality has changed), the second power law clearly 
implies a somewhat faster decay than found in Eq.~(\ref{eq3}). Finally, 
for $H_1 D^{\Delta_1/\nu} \approx 1$ a crossover behavior should be 
detectable, but to clearly identify such behavior may be difficult. In any 
case, the above discussion already shows that (confined) colloid--polymer 
mixtures may give rise to a much more complex behavior than the simple 
Ising model \cite{11}, and hence their study should be rewarding.

\section{Model and simulation technique}

In this work, the colloids are modeled as hard spheres with diameter 
$\sigma_{\rm c}$, while the polymers are described as spheres with 
diameter $\sigma_{\rm p}$. The polymers may not overlap with colloids, but 
there is no interaction between the polymers. The choice of this model 
\cite{23,24} is motivated by the fact that flexible polymers form random 
coils with a rather large gyration radius $R_{\rm g}$, which may 
interpenetrate at very low energy cost (in particular if the solution in 
which the polymers are dissolved, and the colloidal particles are 
suspended, is close to Theta--point conditions \cite{46}). Colloids, on 
the other hand, can be prepared with interactions of very short range 
\cite{16,17,18}.

We specialize to a size ratio $q=\sigma_{\rm p}/\sigma_{\rm c} =0.8$ in 
the following and choose $\sigma_{\rm c} =1$ as our unit of length. The 
polymer and colloid packing fractions are $\eta_{\rm p} = \pi \sigma_{\rm 
p} ^3 N_{\rm p}/(6V)$ and $\eta_{\rm c} = \pi \sigma_{\rm c}^3 N_{\rm 
c}/(6V)$, when we have $N_{\rm p}$ polymers and $N_{\rm c}$ colloids in 
our system of volume $V$, respectively. We choose a volume in the geometry 
of a rectangular box of linear dimensions $L_x \times L_y \times L_z$ with 
$L_x = L_y =L$ and $L_z =D$, respectively. Periodic boundary conditions 
are applied in the $x$ and $y$--directions. In the remaining 
$z$--direction, we place two $L \times L$ walls, at $z=0$ and $z=D$, 
respectively, which act as hard walls for both polymers and colloids, 
while in addition a repulsive step potential of height $\epsilon$ 
(in units of $\kb$) acts on the colloids. The full colloid-wall 
interaction $u_{\rm cw}(h)$ thus reads as:
\begin{equation}\label{eq:cw}
u_{\rm cw}(h) = \left\{
    \begin{array}{ll}
    \infty & \mbox{for $h \leq 0$,} \\
    \epsilon & \mbox{for $0 < h < \sigma_{\rm c}/2$,} \\
    0 & \mbox{otherwise,}
    \end{array}
  \right. \quad
\end{equation}
with $h$ the distance from the {\it surface} of the colloid to the wall. 
In this work, $\epsilon=0$, $0.25$, $0.5$, $1.0$ and $2.0$ are considered. 
Note that the case $\epsilon=0$ (purely hard walls for both colloids and 
polymers) and $D=5$ has already been investigated in our previous work 
\cite{8}. In this case, a very strong attraction between the colloids and 
the walls develops due to the depletion effect \cite{26}.

Following common practice \cite{19,29,30}, we choose the so-called 
(dimensionless) ``polymer reservoir packing fraction'' $\etapr \equiv \pi 
z_{\rm p} \sigma_{\rm p}^3/6$, rather than the polymer fugacity $z_{\rm 
p}$, as the temperature-like variable. As in our previous work 
\cite{8,29,30}, we apply a grand canonical cluster move \cite{29}, 
together with a very efficient reweighting scheme, successive umbrella 
sampling \cite{47}, in order to obtain the distribution function 
$P_L(\eta_{\rm c}|\eta_{\rm p}^{\rm r}, z_{\rm c})$, defined as the 
probability of observing the system with colloid packing fraction 
$\eta_{\rm c}$ at ``inverse temperature'' $\eta_{\rm p}^{\rm r}$ and 
colloid fugacity $z_{\rm c}$. By the subscript $L$, we remind the reader 
that quite generally there will be finite size effects, and a suitable 
extrapolation to the limit $L \rightarrow \infty$, keeping the film 
thickness $D$ fixed, may be required. While for states that are far away 
from phase coexistence, $P_L (\eta_{\rm c}|\eta_{\rm p}^{\rm r}, z_{\rm 
c}$) is a single--peaked function, near two--phase coexistence a 
double--peak structure develops. The precise location of the fugacity 
$z_{\rm c}$ at which two--phase coexistence occurs is obtained using the 
equal--weight rule \cite{48,49}. The positions of the two peaks in 
$P_L(\eta_{\rm c}|\eta_{\rm p}^{\rm r}, z_{\rm c})$ then yield estimates 
for the two branches, $\etacl$ and $\etacv$, of the binodal. In 
addition, we also study reduced moments of $P_L(\eta_{\rm c}|\eta_{\rm 
p}^{\rm r},z_{\rm c}$) at phase coexistence, defining an analogue of the 
order parameter of the Ising model,
\begin{equation}\label{eq5}
m = \eta_{\rm c} - \langle \eta_{\rm c} \rangle, \; 
\langle \eta_{\rm c}\rangle =
\int \limits_0^\infty \eta_{\rm c} 
P_L (\eta_{\rm c}| \eta_{\rm p}^{\rm r}, z_{\rm c}) d\eta_{\rm c}
\end{equation}
and higher-order moments
\begin{equation}\label{eq6}
\langle m^p\rangle = \int \limits_0^\infty m^p 
P_L (\eta_{\rm c}|\eta_{\rm p}^{\rm r}, z_{\rm c})d\eta_{\rm c} \;.
\end{equation}
As is well known, following the behavior of moment ratios such as
\begin{equation}\label{eq7}
  U_4=\langle m^2\rangle^2 / \langle m^4\rangle
\end{equation}
along the path where $z_{\rm c}$ is at phase coexistence in the $(z_{\rm 
c},\etapr)$-plane, is useful for locating the critical point of the 
system; this is the so-called ``cumulant intersection method'' 
\cite{50,51,52}. Although it has been demonstrated recently \cite{8,53} 
that a study of the full distribution $P_L(\eta_{\rm c}|\eta_{\rm p}^{\rm 
r},z_{\rm c})$ and its moments also for fugacities $z_{\rm c}$ off 
coexistence near criticality is useful, and can yield even more accurate 
information on the critical behavior, the computer time resources for such 
a study in the present case are very demanding, and hence not attempted 
here.

\section{Results for $\epsilon=0.5$}

In this section, we present results using step-height $\epsilon=0.5$ in the
colloid--wall interaction of \eq{eq:cw}. We ultimately aim to test the
Fisher--Nakanishi scaling predictions \cite{5} for the shift in the critical
point parameters. To this end, the location of the critical point as function of
the film thickness $D$, the binodal, and the behavior of the critical colloid
packing fraction are investigated first.

\begin{figure}
\centering
\includegraphics*[width=.4\textwidth]{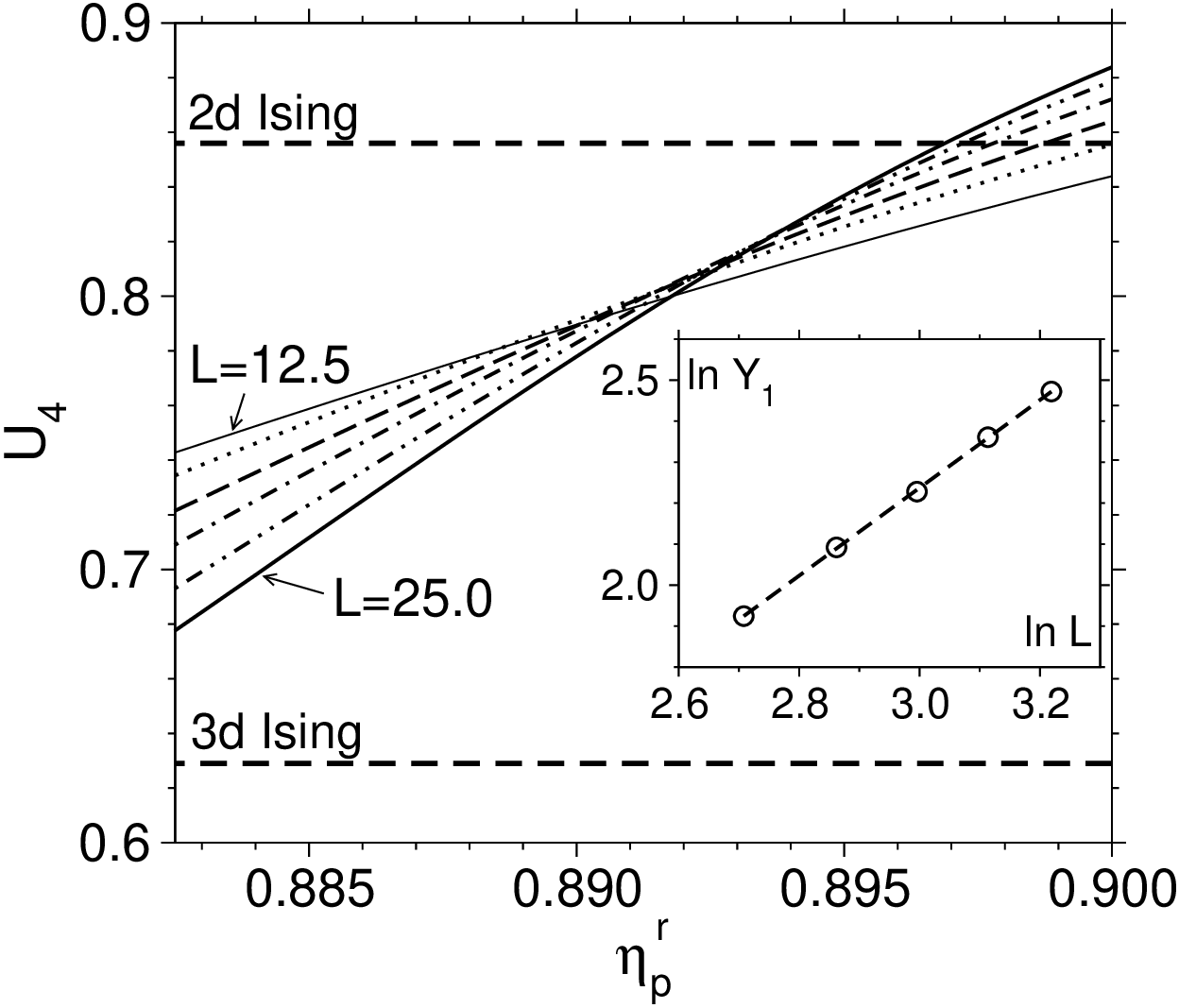}
\caption{\label{fig1}Moment ratio $U_4$ for a film of thickness $D=5$ 
plotted versus the polymer reservoir packing fraction $\eta_{\rm p}^{\rm 
r}$ and several choices of the lateral dimension $L=12.5$, 15.0, 17.5, 
20.0, 22.5, and 25.0. The intersection point yields an estimate for the 
critical ``inverse temperature'' $\etaprcr = 0.892$. The inset shows a 
log--log plot of the cumulant slope $Y_1$ versus the linear dimension 
$L$. The straight line yields the exponent $1/\nu _{\rm eff}$, from which 
we deduce $\nu_{\rm eff}=0.933$.}
\end{figure}

\begin{figure}
\centering
\includegraphics*[width=.4\textwidth]{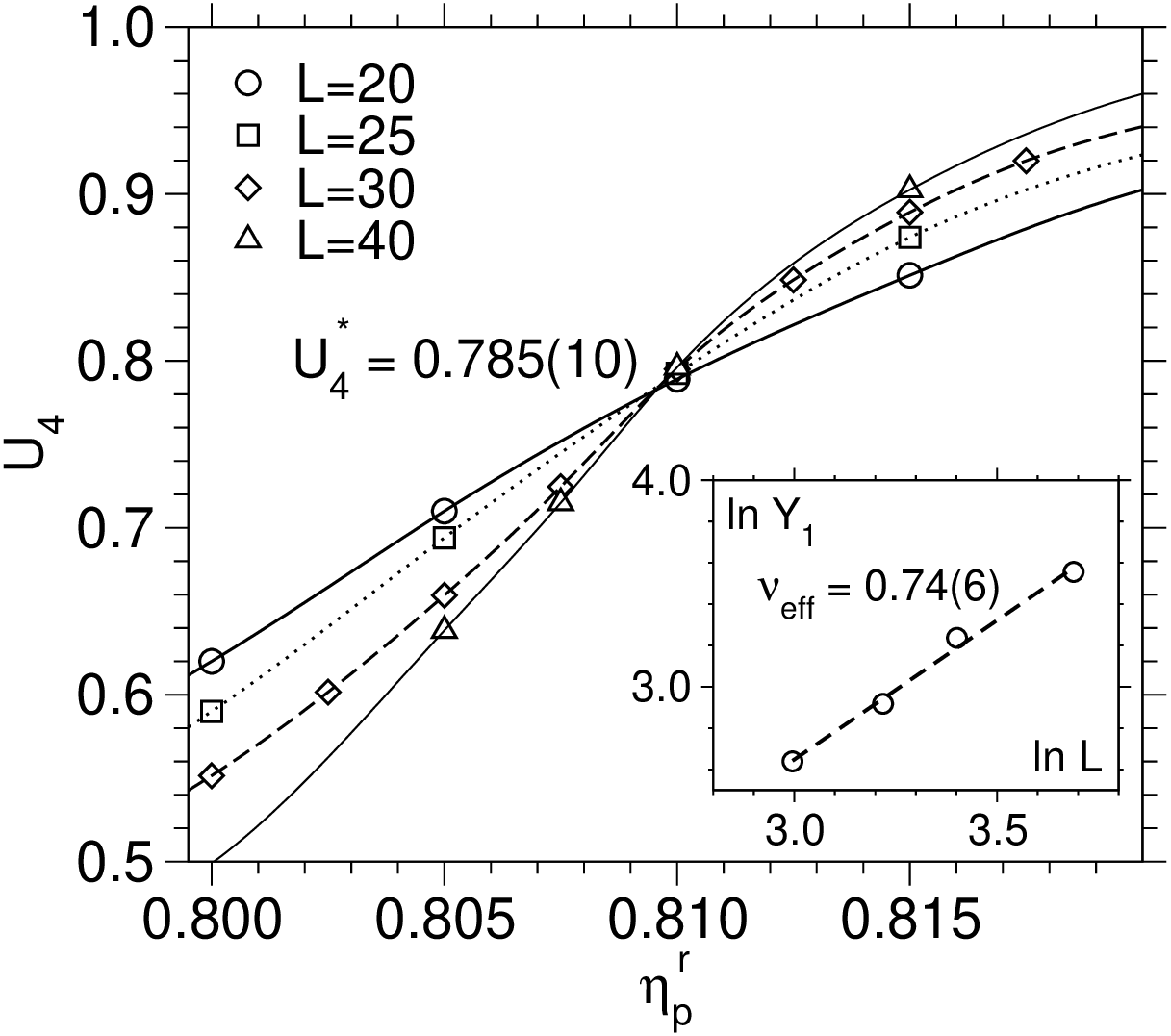}
\caption{Same as Fig.~\ref{fig1}, but for film thickness $D=10$, yielding 
the estimates $\etaprcr = 0.810$ and $\nu_{\rm eff} = 0.74$. \label{fig2}}
\end{figure}

\begin{figure}
\centering
\includegraphics*[width=.4\textwidth]{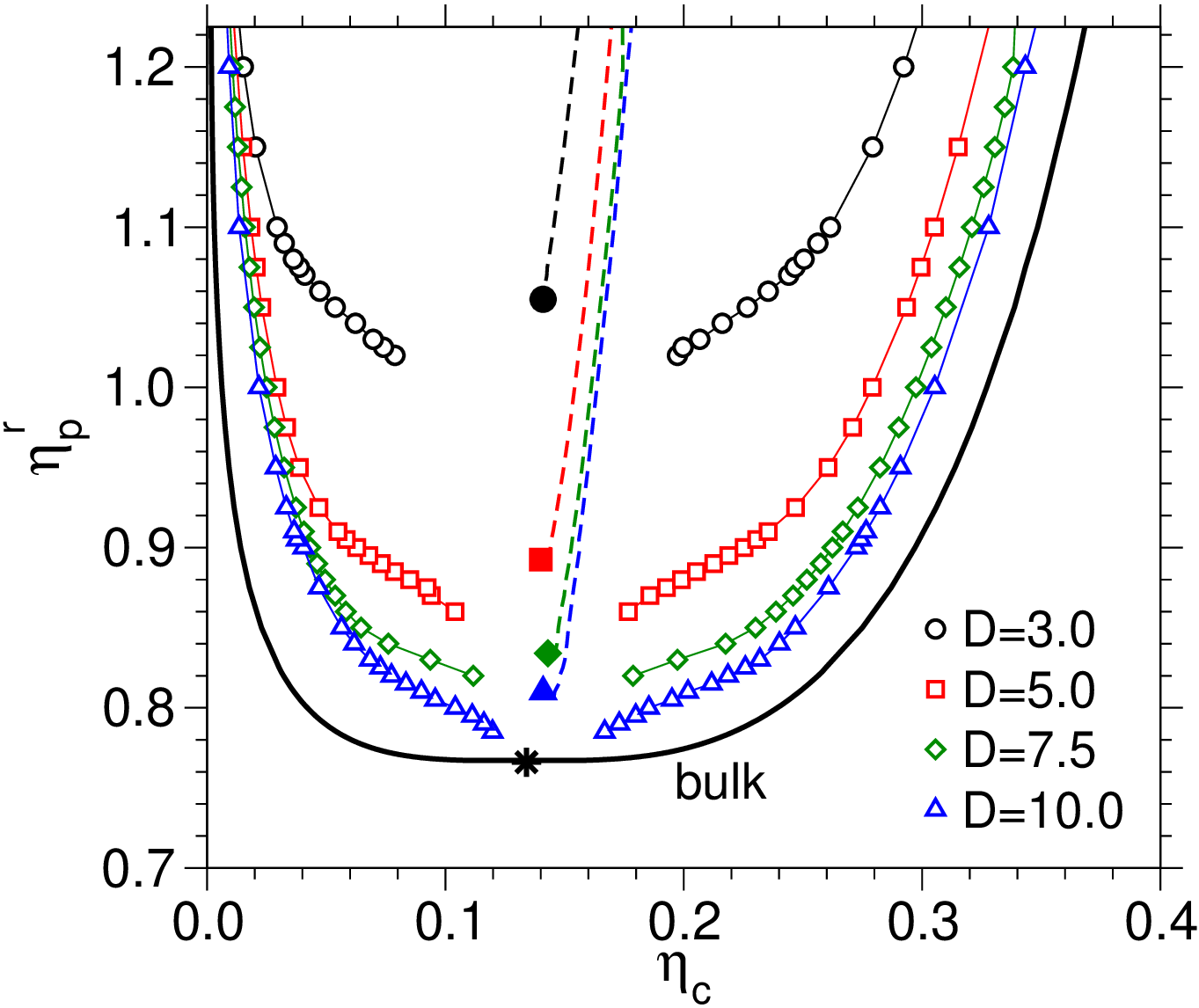}
\caption{Binodals of the AO--model with $q=0.8$ in bulk (full curve without data
points) and in confinement by parallel hard walls, to which a repulsive
potential on the colloids with strength $\epsilon = 0.5$ was added (see
\eq{eq:cw}). Circles denote data for $D=3$, $L=18$; squares for $D=5$, $L=20$;
diamonds for $D=75$, $L=30$; and triangles for $D=10$, $L=30$. The dotted lines
denote the estimated coexistence diameters.\label{fig3}} 
\end{figure}

\begin{table}
\begin{center}
\begin{tabular}{|c|c|c|c|c|c|}
\hline $D$ & $\nu$ & $U_4^*$ & $\etaprcr$ & $\mu_{\rm c,cr}^{\rm coex}(D)$ & $\delta^*$  \\ \hline \hline
2d bulk & 1        & 0.856     & ---      & ---       & ---      \\ \hline
3.0     & 0.955(5) & 0.84(1)   & 1.055(5) & 4.39(2)   & 0.139(1) \\ \hline
5.0     & 0.933(3) & 0.81(1)   & 0.892(4) & 3.715(15) & 0.142(1) \\ \hline
7.5     & 0.805(5) & 0.80(1)   & 0.834(4) & 3.43(2)   & 0.143(1) \\ \hline
10.0    & 0.74(6)  & 0.785(10) & 0.810(5) & 3.307(7)  & 0.141(1) \\ \hline
3d bulk & 0.630    & 0.629     & 0.766    & 3.063     & 0.134    \\ \hline
\end{tabular}
\end{center}
\caption{\label{table1} Critical parameters of the AO model for different film
thicknesses $D$, compared to the corresponding 2d and 3d bulk values. Note that
the simulation data reported in this table were obtained using $\epsilon=0.5$ as
step-height in the colloid--wall interaction of \eq{eq:cw}.}
\end{table}

\subsection{Critical point}

We aim to locate the critical value of $\etapr$, for a number of 
thicknesses $D$. To this end, the cumulant ratio $U_4$, see \eq{eq7}, was 
measured as function of $\etapr$ using several lateral dimensions $L$. The 
results are shown in Figs.~\ref{fig1} and \ref{fig2}, for $D=5$ and 
$D=10$, respectively. We found that for $D \leq 10$, simulation data of 
meaningful accuracy could still be generated, although for $D=10$ there is 
the need to choose the lateral linear dimension $L$ as large as $L=40$. 
%This implies, at the critical point, a total of about $N_{\rm c} \approx 
%XXX$ colloidal particles, and $N_{\rm p} \approx XXX$ polymers, which 
%significantly exceeds the system size reached in our previous work. 
From the cumulant intersection points, $\etaprcr$ is obtained. The 
``effective'' correlation length exponent $\nu_{\rm eff}$ is obtained from 
the $L$-dependence of the cumulant slope $Y_1$ evaluated at the 
intersection point. One expects that $Y_1 \propto L^{1/\nu_{\rm eff}}$; 
the insets in Figs.~\ref{fig1} and \ref{fig2} show that our simulation 
estimates of $Y_1$ are indeed compatible with this relation. Additional 
simulations were performed for thicknesses $D=3$ and $D=7.5$; the 
resulting estimates of $\etaprcr$ and $\nu_{\rm eff}$, as well as the 
cumulant value $U_4^*$ at the critical point, and the corresponding 
coexistence colloid chemical potential, are collected in 
Table~\ref{table1}. From these data, it is clear that, by increasing the 
film thickness $D$, a gradual crossover in the effective critical behavior 
from 2d~Ising toward 3d~Ising universality occurs. While for $D=3$, 
$\nu_{\rm eff}$ is very close to the exact value $\nu=1$ of the 2d~Ising 
model, and similarly for the corresponding cumulant value $U_4^*$ (for the 
2d~Ising model, the very precise estimate $U^*_4=0.856$ was obtained 
\cite{54}), with increasing $D$ a clear shift toward 3d~Ising values is 
observed.

Of course, the smooth decrease of these ``effective'' values can only
be taken as a very rough description of the theoretically expected
crossover scaling scenario \cite{8,9,10,11}. For the case of $D=5$ and
hard-wall boundary conditions, a more elaborate analysis is presented
in \olcite{8}.  While in this case the cumulant slopes yield the same
``effective'' exponent $\nu_{\rm eff} \approx 0.933$ as found here for
$D=5$ \cite{8}, the more elaborate analysis in fact revealed \cite{8}
that the asymptotic critical behavior is in reality described {\it
precisely} by 2d~critical exponents, but this can be seen only in a very
narrow region around the critical point. Analyzing any such critical
quantity on a log--log plot, one typically finds a systematic curvature:
the slope of the curve approaches the 2d value for very small $|t|$ and
then very gradually bends over in the direction towards the 3d value
(the latter is not really observed because the crossover is not yet
complete as $|t|$ becomes so large that one leaves the critical region)
\cite{8}. Since a thorough analysis of crossover scaling requires an
enormous investment of computer resources, we have refrained from doing
so, since there is no reason to expect any significant surprises.

\subsection{Binodal and coexistence diameter}

Next, we consider the binodal. Fig.~\ref{fig3} shows some of our ``raw'' data
for the binodals in confinement, together with our estimates of the
corresponding coexistence diameters (broken lines). By ``raw'' we mean that no
finite size scaling analysis was performed on the data. The coexistence diameter
$\delta$ is defined as the average of the coexisting phase densities
\begin{equation}\label{eq8}
  \delta = (\etacl + \etacv)/2,
\end{equation}
where, as before, $\etacv$ and $\etacl$ represent the coexistence colloid
packing fractions on the vapor and liquid branches of the binodal,
respectively. The diameters terminate at the critical value of $\etapr$,
for which we used the estimates listed in Table~\ref{table1}. Since
the data were obtained for finite $L \leq 30$, the two branches of the
binodal do not merge at the critical point, but rather extend beyond
it, bending over into the one--phase--region.  This effect of ``finite
size tails'' or ``rounding'' of the order parameter into the disordered
region of the phase diagram is well--known from simulations of the Ising
model \cite{50,51,52}. It is due to the fact that the order parameter
distribution is clearly double peaked right at criticality and also in
the one--phase region, if the lateral linear dimension $L$ does not
yet exceed the correlation length \cite{50,51,52}.  The exception in
Fig.~\ref{fig3} is the full curve, which represents the binodal of the
3d~bulk AO model. This curve was obtained using the finite size scaling
approach of \olcite{53}, and, on the scale of \fig{fig3}, should rather
precisely reflect the true thermodynamic limit form $L \to \infty$
of the bulk binodal.

\begin{figure}
\centering
\includegraphics*[width=.4\textwidth]{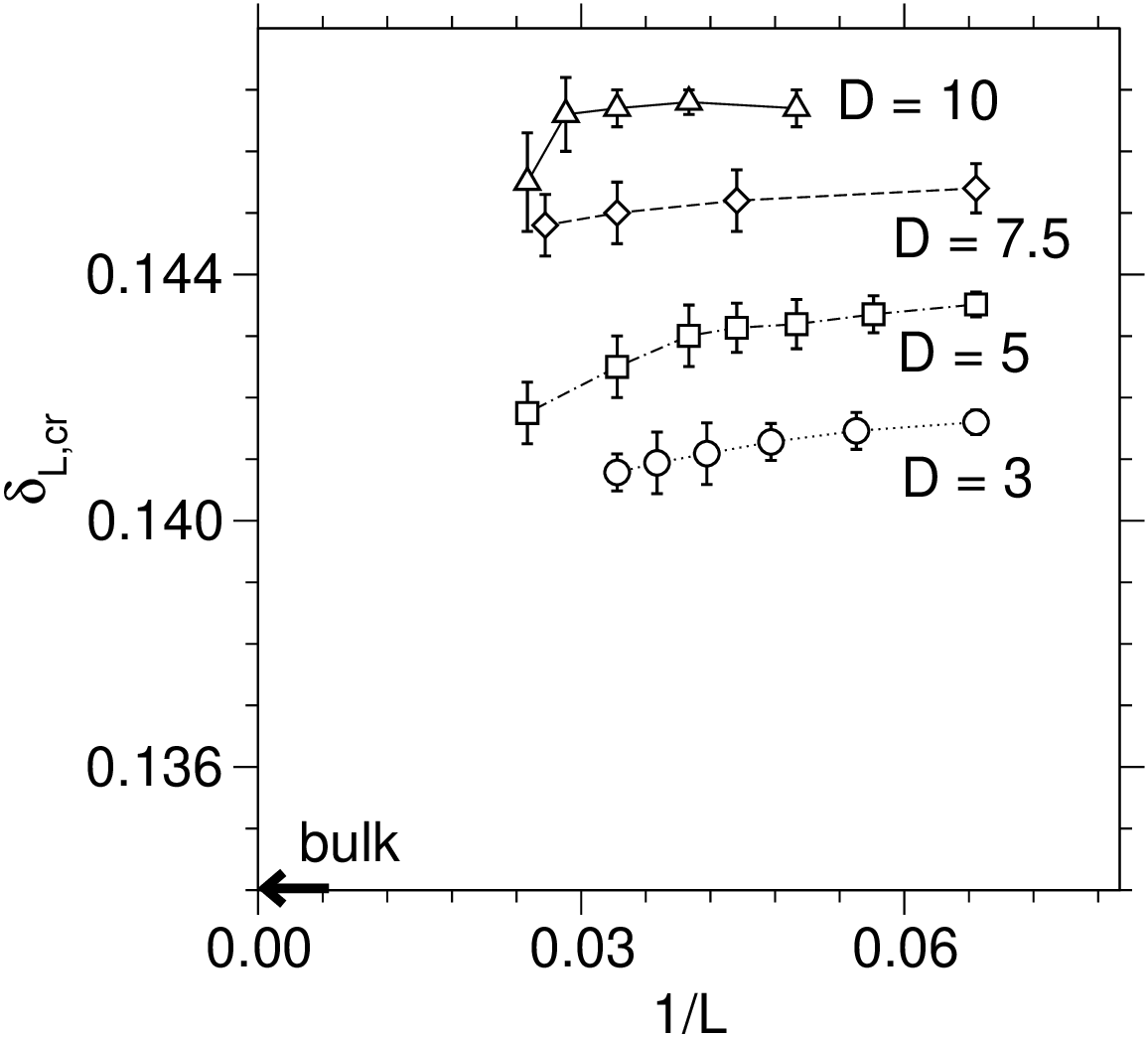}
\caption{\label{fig4} Finite size coexistence diameters $\delta_{L,\rm cr}$ as
function of $1/L$, for various choices of $D$ as indicated. The arrow marks the
diameter at the critical point in the thermodynamic limit of the bulk system,
i.e.~in the absence of walls, and was taken from \olcite{30}.}
\end{figure}

\begin{figure}
\centering
\includegraphics*[width=.4\textwidth]{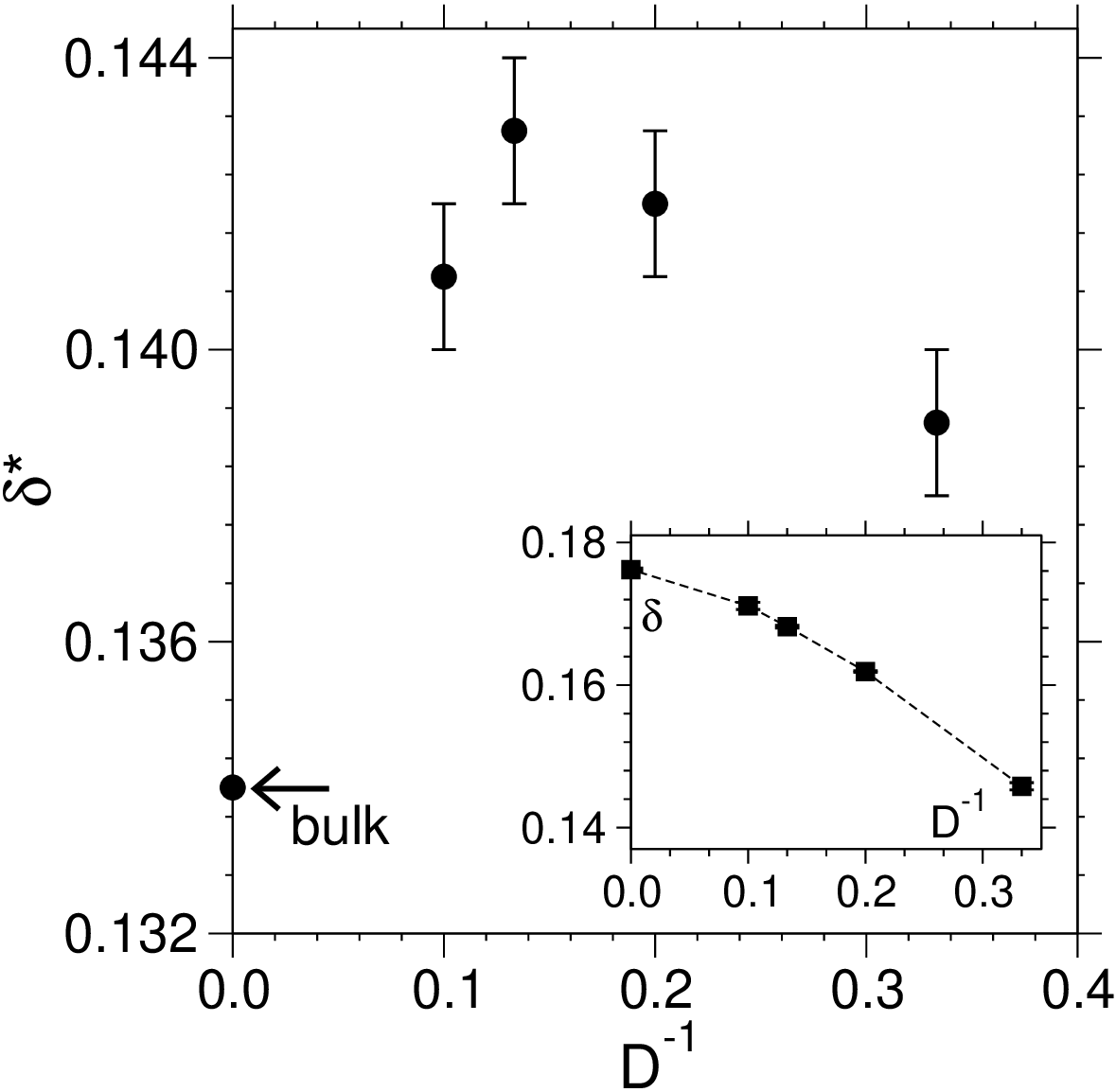}
\caption{\label{fig5} Variation of the diameter at the critical point 
$\delta^* = \lim_{L \rightarrow \infty} \delta_{L,\rm cr}$ as function of 
$D^{-1}$. The inset shows the diameter $\delta$ away from the critical 
point, choosing $\etapr=1.1$, as function of $D^{-1}$.}
\end{figure}

In order to more accurately determine the colloid packing fraction at the 
critical point $\eta_{\rm c,cr}$, we have examined the $L$-dependence of 
$\delta_{L,\rm cr}$, defined as the value of the coexistence diameter 
$\delta$, given by \eq{eq8}, as obtained in a finite system of lateral 
dimension $L$, at the critical value of $\etapr$. The result is shown in 
Fig.~\ref{fig4}, where $\delta_{L,\rm cr}$ is plotted as function of 
$1/L$, using for $\etaprcr$ the values listed in Table~\ref{table1}. The 
data display significant scatter, but rather intriguing behavior is 
revealed nevertheless. By increasing the film thickness $D$, also 
$\delta_{L,\rm cr}$ increases, away from the bulk value (arrow in 
Fig.~\ref{fig4}), although in the limit $D \to \infty$ precisely this bulk 
value should be recovered again. A possible explanation for this 
non-monotonic behavior with $D$, may be found in the precise critical 
behavior of the coexistence diameter in the thermodynamic limit. The 
critical behavior of $\delta$ in the limit $L \rightarrow \infty$ is 
rather intricate, and governed by several competing singular terms 
\cite{55}
\begin{equation}\label{eq9}
  \delta = \eta_{\rm c,cr} \left(1 + A_{2\beta} t^{2\beta} + A_{1-\alpha}
  t^{1-\alpha} + A_1 t + \ldots \right),
\end{equation}
with relative distance from the critical point $t \equiv 
\etapr/\etaprcr-1$, (nonuniversal) amplitudes $A_i$, and $\etaccr$ the 
colloid packing fraction at the critical point. Note that in 2d, the first 
singular term has a much smaller exponent $(2\beta=1/4$), but also the 
amplitude $A_{2\beta}$ may be very small compared to the amplitudes of the 
other terms. The next order term may involve a logarithmic correction 
\cite{56}; recall that $\alpha=0$ in the 2d Ising model implies a 
logarithmic divergence of the specific heat. It is expected that in a 
finite system, the singular terms $t^{2\beta} \propto \xi^{-2 \beta/\nu}$ 
and $t^{1-\alpha} \propto \xi^{-(1-\alpha)/\nu}$, cross--over to 
correction terms, with $\xi$ ultimately replaced by $L$. Hence, we expect 
for the quantity $\delta_{L,\rm cr}$ in Fig.~\ref{fig4} the scaling form
\begin{eqnarray}\label{eq10}
  \delta_{L,\rm cr} &=& \eta_{\rm c,cr}(D) [
  1 + \hat{A}_{2\beta}(D/L) L^{-2\beta/\nu} \nonumber \\
  &+& \hat{A}_{1-\alpha}(D/L) L^{-(1-\alpha)/\nu} \nonumber \\
  &+& \hat{A}_1(D/L) L^{-1/\nu} + \cdots ].
\end{eqnarray}
Here, we have assumed that the amplitudes $\hat{A}_i$ are functions of the 
aspect ratio $D/L$; such a dependence can be motivated by finite size 
scaling arguments \cite{10,11}. In view of the rather large error bars in 
$\delta_{L,\rm cr}$ in Fig.~\ref{fig4}, and the complicated structure of 
\eq{eq10}, we feel that an accurate extrapolation to obtain $\etaccr(D)$ 
is not possible. Similarly, already in \olcite{8}, it was pointed out that 
a reliable estimate of the critical behavior of the diameter in 
confinement from simulation data is not yet feasible. Hence, the only 
tentative conclusion we can draw from Fig.~\ref{fig4} is that presumably 
the dependence of $\eta_{\rm c,cr}(D)$ on $D$ is non-monotonic: $\eta_{\rm 
c,cr}(D)$ for small $D$ does not differ much from $\eta_{\rm 
c,cr}(\infty)$, but with increasing $D$ the difference increases first, 
reaches a maximum, and then decreases again. Such behavior could stem from 
competing signs of some of the amplitudes in Eqs.~(\ref{eq9}) 
and~(\ref{eq10}).

For completeness, Fig.~\ref{fig5} shows $\delta^* \equiv \lim_{L \to \infty}
\delta_{L,\rm cr}$ as function of $D^{-1}$, where $\delta^*$ was obtained by
linearly extrapolating the data for $\delta_{L,\rm cr}$ of Fig.~\ref{fig4} in
the variable $1/L$. The resulting estimates of $\delta^*$ have also been
collected in Table~\ref{table1}, and reflect our best possible estimates of the
critical colloid packing packing fraction $\etaccr$ in the thermodynamic limit
(recall that \eq{eq9} for the diameter reduces to $\etaccr = \delta^*$ at the
critical point $t \to 0$). It is plausible from these data that the variation of
$\delta^*$ with $D^{-1}$ is indeed non-monotonic. However, the large error bars
in $\delta^*$, resulting from the uncertainty in the extrapolation of
$\delta_{L,\rm cr}$ to $L \to \infty$, prevent us from making any quantitative
statements. In contrast, at values of $\etapr$ that are far from the critical
point for all choices of $D$ considered here, such as $\etapr=1.1$, the
$L$-dependence in the diameter can safely be neglected. In this case, the
variation with $D^{-1}$ is clearly monotonic, and approximately linear for small
$D^{-1}$, see the inset of Fig.~\ref{fig5}.

\begin{figure}
\centering
\includegraphics*[width=.45\textwidth]{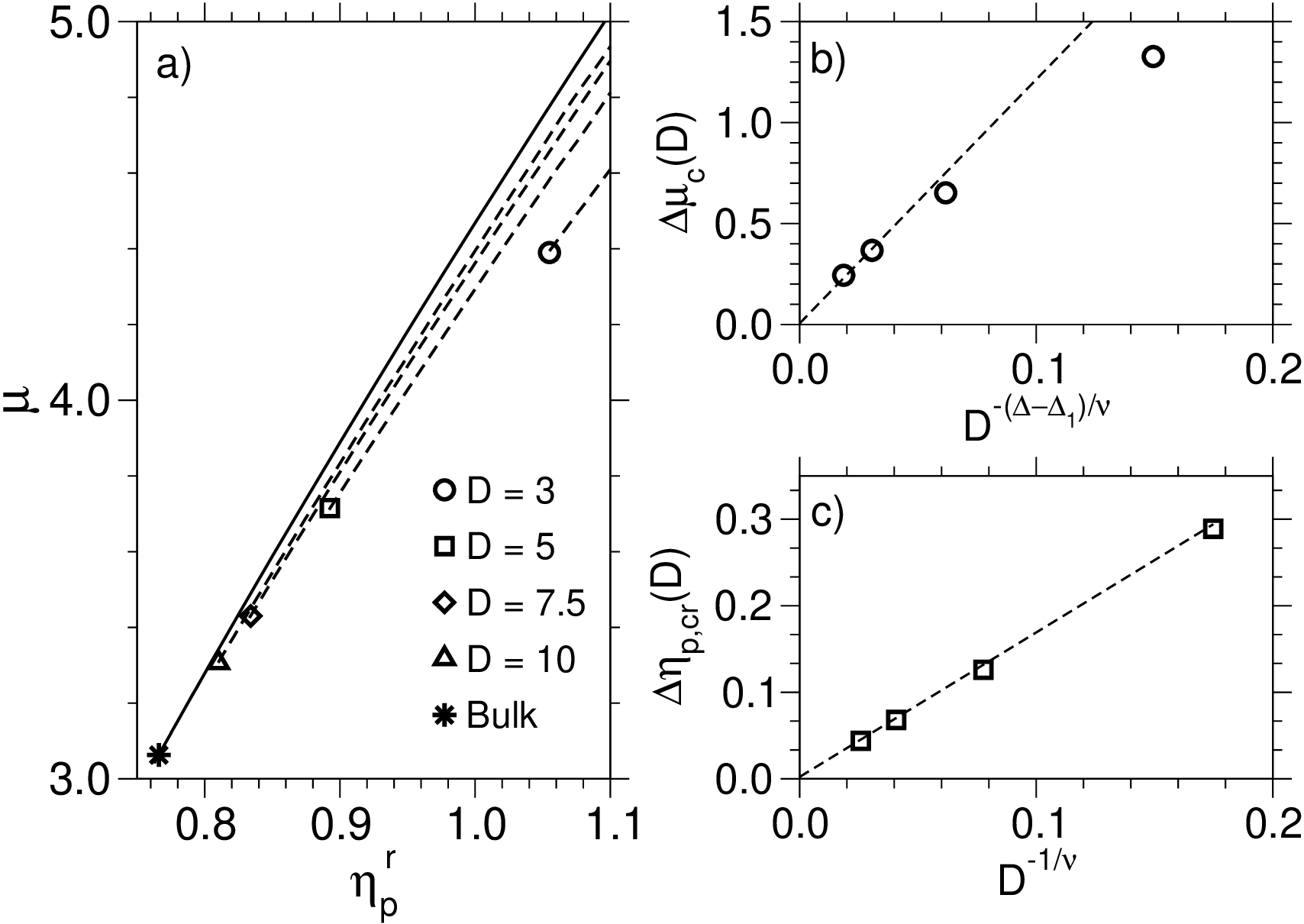}

\caption{\label{fig6} (a) Binodals of the AO-model with $q=0.8$ in 
so-called grand-canonical representation, choosing the coexistence 
chemical potential of the colloids $\mu_{\rm c}^{\rm coex}$ and $\etapr$ 
as variables. Shown is the bulk binodal (full curve), and the binodal in 
confinement for several film thicknesses $D$ (broken curves). The symbols 
mark the corresponding critical points. (b) Shift of the critical 
coexistence colloid chemical potential, given by \eq{eq11a}, as function 
of $D^{-(\Delta-\Delta_1)/\nu}$. (c) Shift of the critical polymer 
reservoir packing fraction, given by \eq{eq11b}, as function of 
$D^{-1/\nu}$. Linear straight line fits through the origin in (b) and (c) 
confirm that the data are compatible with the theoretical 
Fisher--Nakanishi predictions \cite{5}.}

\end{figure}

\subsection{Fisher--Nakanishi scaling and Kelvin equation}

Next, we consider the Fisher--Nakanishi scaling predictions, essentially 
the main result of this paper. Recall from the discussion in Section~II, 
that the Fisher--Nakanishi scaling predictions pertain to the shift in the 
critical point parameters. More precisely, for the shift in the critical 
``inverse temperature'' as function of the film thickness $D$, one expects 
that
\begin{equation}\label{eq11b}
  \Delta\etaprcr(D) \equiv \etaprcr(D) - \etaprcr(\infty) \propto 
D^{-1/\nu},
\end{equation}
with $\nu$ the correlation length critical exponent of the 3d~Ising model,
$\eta_{\rm p, cr}^{\rm r}(D)$ the critical value of $\eta_{\rm p}^{\rm
r}$ in the confined system of thickness $D$, and $\eta_{\rm p, cr}^{\rm
r}(\infty)$ the critical value of $\etapr$ in the bulk ($D \to \infty$)
system. The second scaling prediction of Fisher and Nakanishi pertains
to the shift in the coexistence chemical potential of the colloids at
criticality, as function of the film thickness $D$. It is expected that
\begin{equation}\label{eq11a}
 \Delta\mu_{\rm c,cr}^{\rm coex}(D) \equiv \mu_{\rm c,cr}^{\rm coex}(D) - 
 \mu_{\rm c,cr}^{\rm coex}(\infty) \propto D^{(\Delta_1-\Delta)/\nu},
\end{equation}
with $\Delta$ and $\Delta_1$ the gap exponents introduced in Section~II, 
$\nu$ again the correlation length exponent of the 3d~Ising model, 
$\mu_{\rm c,cr}^{\rm coex}(D)$ the coexistence chemical potential of the 
colloids at criticality in the confined system, and $\mu_{\rm c,cr}^{\rm 
coex}(\infty)$ the coexistence chemical potential of the colloids at 
criticality in the bulk system. Our results have been collected in 
Fig.~\ref{fig6}. The left panel shows the coexistence chemical potential 
$\mu_{\rm c}^{\rm coex}$ of the colloids as function of $\etapr$, for 
several values of the film thickness $D$. We remind the reader that the 
coexistence chemical potential of the colloids follows from the 
equal--weight--rule \cite{48,49}. The shift in the coexistence chemical 
potential of the colloids at criticality, see \eq{eq11a}, is plotted in 
\fig{fig6}(b), as function of $D^{(\Delta_1-\Delta)/\nu}$. Here, the 
chemical potential shifts were simply read-off from \fig{fig6}(a); the 
corresponding values have also been collected in Table~\ref{table1} for 
completeness. The shift in the critical ``inverse temperature'', see 
\eq{eq11b}, is plotted in \fig{fig6}(c), as function of $D^{-1/\nu}$, 
where for the critical values of $\etapr$, the estimates of Table~\ref{table1} 
were used. In Figs.~\ref{fig6}(b) and (c), the known values of the 
exponents were used, as quoted previously. Validity of the 
Fisher--Nakanishi relations implies that the data collapse onto straight 
lines through the origin. Of course, data for small $D$, such as $D=3$ and 
$D=5$, need not follow these relations, because the Fisher--Nakanishi 
predictions are valid only for {\it asymptotically} large $D$. For small 
$D$, corrections to finite size scaling likely come into play. In view of 
this, the agreement of our data with the scaling predictions, 
Eqs.~(\ref{eq2}) and (\ref{eq3}), is already rather satisfactory.

\begin{figure}
\centering
\includegraphics*[width=.4\textwidth]{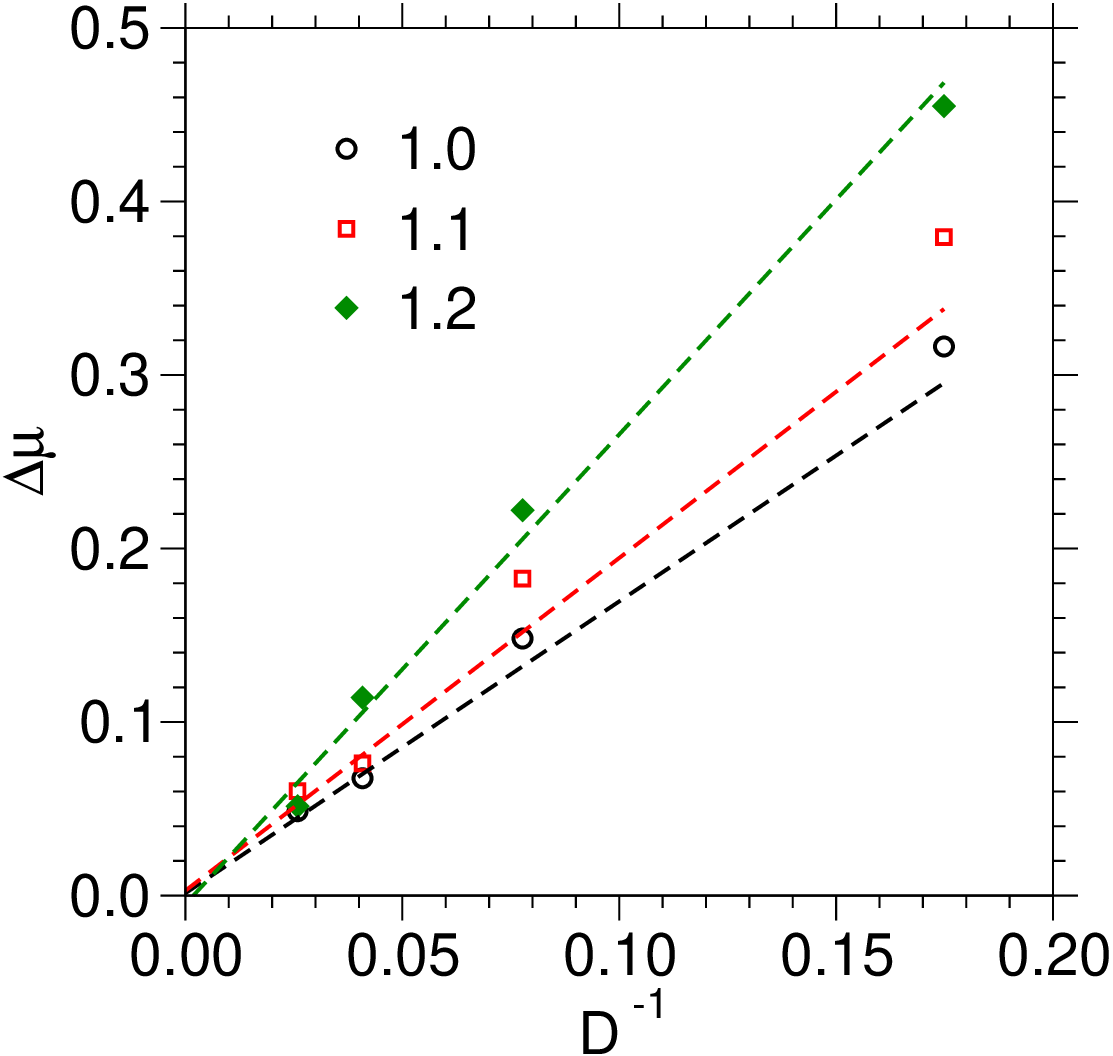}
\caption{\label{fig7} Test of the Kelvin equation. Shown is the chemical 
potential difference, given by \eq{eq:kelvin}, as function of the inverse 
film thickness. Three values of $\etapr$, chosen well above the critical 
values $\etaprcr(D)$, are included. Broken straight lines show that the 
data are compatible with the Kelvin equation.} 
\end{figure}

Finally, we consider the Kelvin equation, which is expected to describe 
the shift in the coexistence colloid chemical potential away from the 
critical point. It is expected that
\begin{equation}\label{eq:kelvin}
  (\Delta\mu)_{\rm Kelvin} \equiv \mu_{\rm c}^{\rm coex}(D)  - 
  \mu_{\rm c}^{\rm coex}(\infty) \propto 1/D,
\end{equation}
with $\mu_{\rm c}^{\rm coex}(D)$ the coexistence chemical potential of the 
colloids in the confined system, and $\mu_{\rm c}^{\rm coex}(\infty)$ that 
in the bulk. Fig.~\ref{fig7} presents a test of the Kelvin equation, 
plotting the chemical potential difference as function of $D^{-1}$, 
choosing three values of $\etapr$ which are much {\it higher} than the 
corresponding critical values $\etaprcr(D)$. Once again, some systematic 
deviations at very small thicknesses ($D=3$ and $D=5$) are seen, while for 
$D=7.5$ and $D=10$, one can already observe the expected straight lines 
through the origin.

\subsection{Structural properties}

\begin{figure}
\centering
\includegraphics*[width=.4\textwidth]{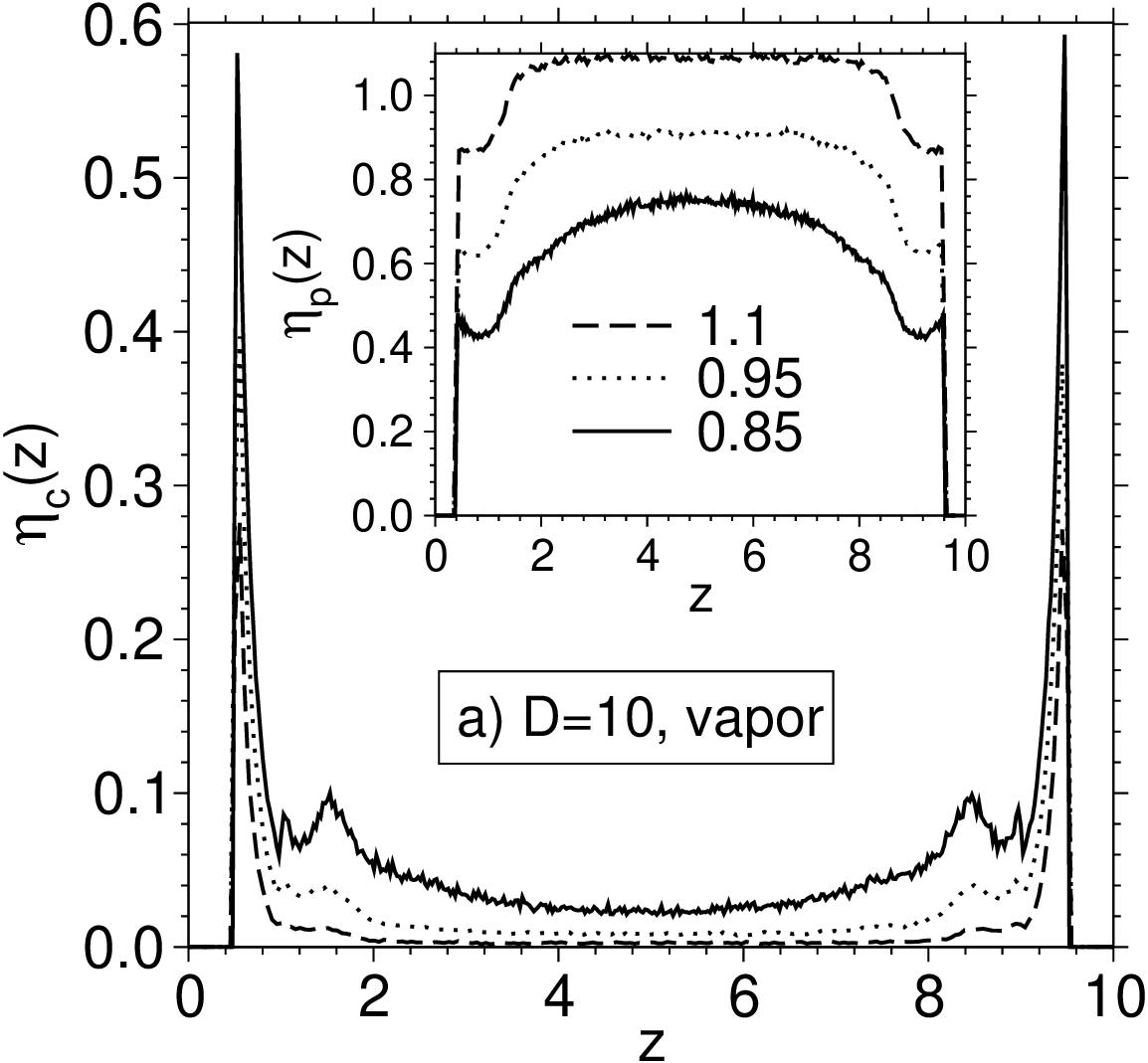}
\includegraphics*[width=.4\textwidth]{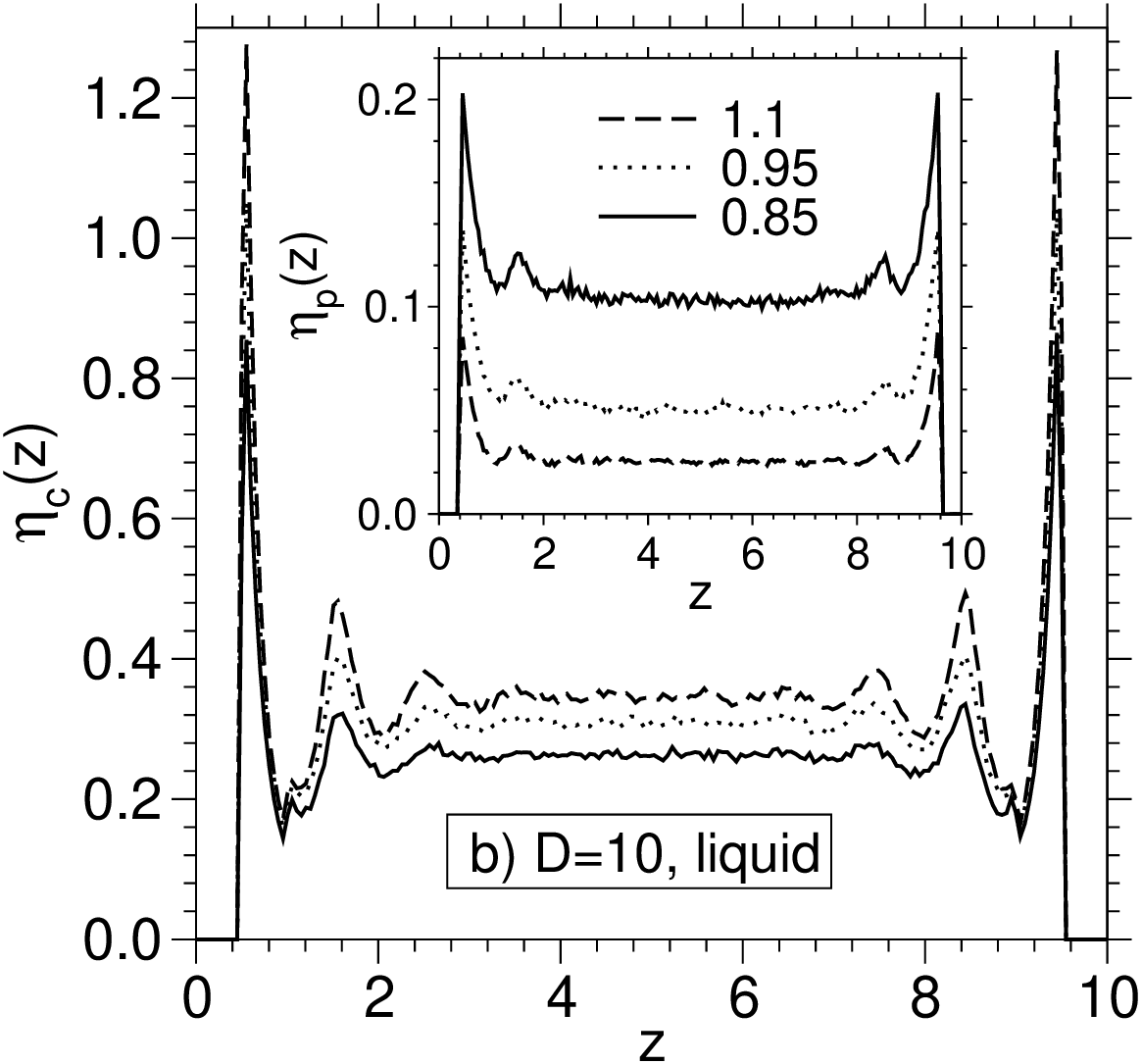}
\caption{\label{fig8} Colloid density profiles for film thickness $D=10$ 
and lateral dimension $L=30$. Profiles were obtained for three different 
values of $\etapr$, as indicated, chosen to be well away from the critical 
point. Shown are profiles measured on the vapor branch of the binodal (a) 
and on the liquid branch (b). The insets show the corresponding polymer 
density profiles.}
\end{figure}

\begin{figure}
\centering
\includegraphics*[width=.4\textwidth]{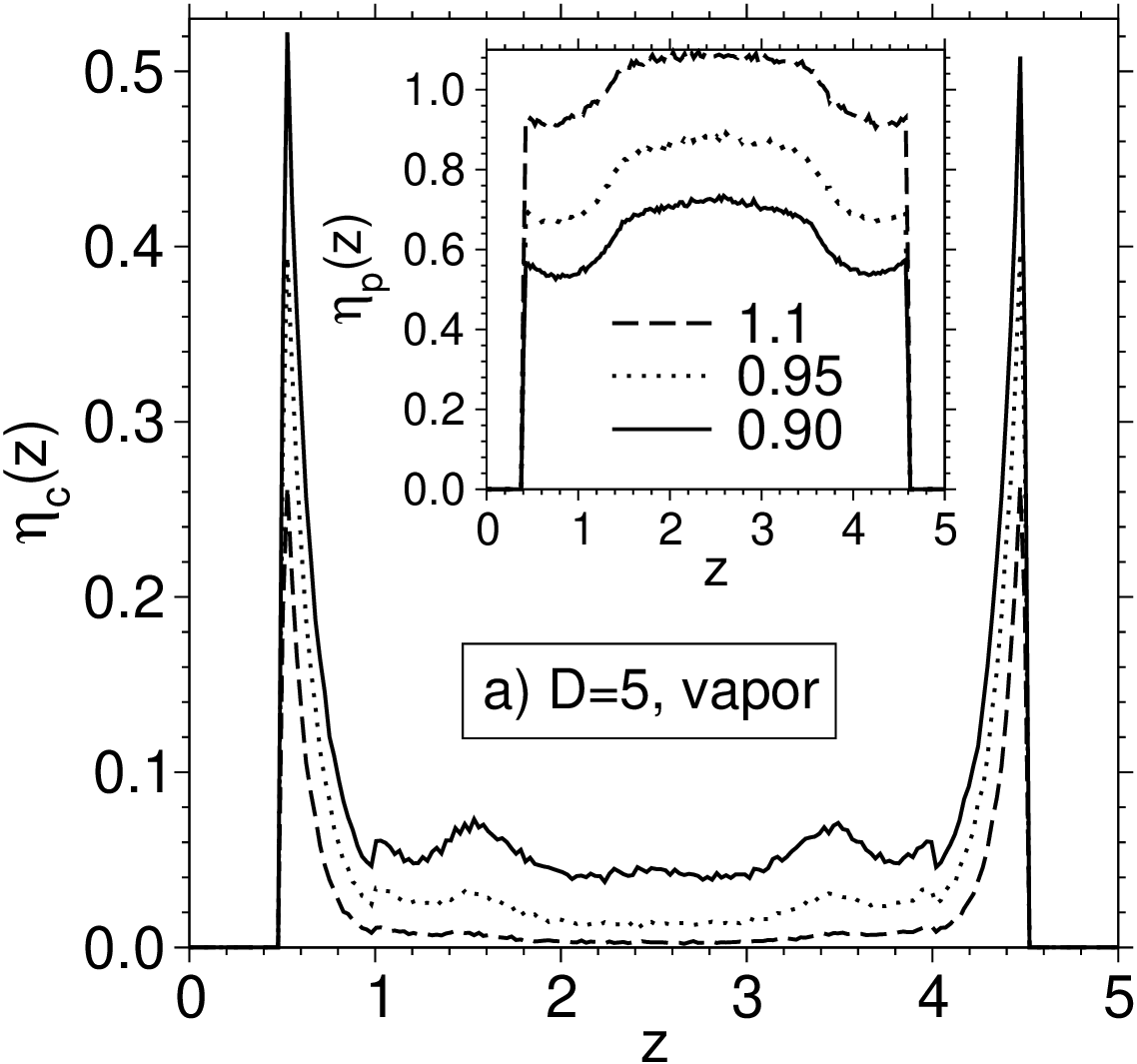}
\includegraphics*[width=.4\textwidth]{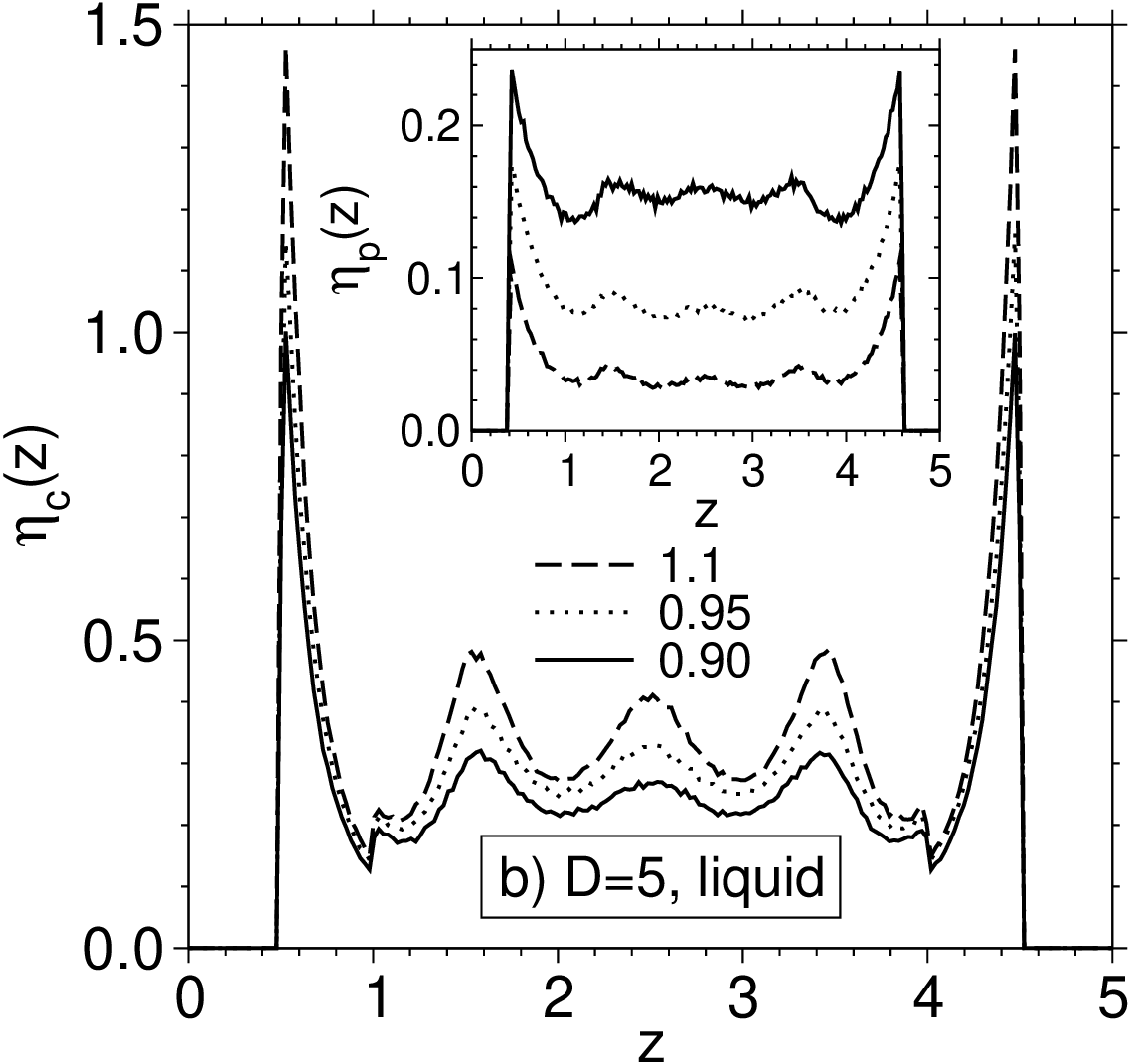}
\caption{\label{fig9} Same as \fig{fig8}, but for $D=5$ and $L=20$.}
\end{figure}

We now consider the local structure in the thin films. Figs.~\ref{fig8} 
and \ref{fig9}, show density profiles of colloids and polymers (the latter 
are shown as insets), as a function of the distance $z$ across the film. 
Results are shown for for $D=10$ (\fig{fig8}) and $D=5$ (\fig{fig9}). The 
three values of $\etapr$ are chosen well away from the critical point, 
such that finite size effects in $L$ can safely be neglected. The upper 
panel in each figure shows profiles obtained in the colloidal ``vapor'' 
phase; the lower panel in the colloidal ``liquid'' phase, that coexists 
with the vapor. It is clear that, despite the repulsive colloid-wall 
energy ($\epsilon=0.5$ in \eq{eq:cw}), the entropically driven attraction 
of the colloids to the walls still exists, and causes the formation of a 
rather dense layer of colloids at the walls. At not too large values of 
$\etapr$, one can even see an indication of a second layer of colloids in 
the ``vapor'' phase. Of course, in the liquid phase, rather pronounced 
layering effects can always be detected. For the case $D=10$, 
there are three colloidal layers adjacent to each wall; in the range $3.5 
\leq z \leq 6.5$, the density profiles are essentially horizontal. This 
flatness indicates bulk-like behavior in the center of the thin film. In 
contrast, for the case $D=5$, the layering associated with the left wall 
``interferes'' with layering effects at the right wall, and no bulk region 
can be identified. For the particular parameters considered here, 
\fig{fig9} shows that only five layers of colloids can be accommodated in 
a slit $D=5$ colloid diameters wide. Note also the step-like increase in 
the colloid density at $z=1$ and $z=4$. This feature must be attributed to 
the step in the colloid-wall interaction, see \eq{eq:cw}.

\begin{figure}
\centering
\includegraphics*[width=.4\textwidth]{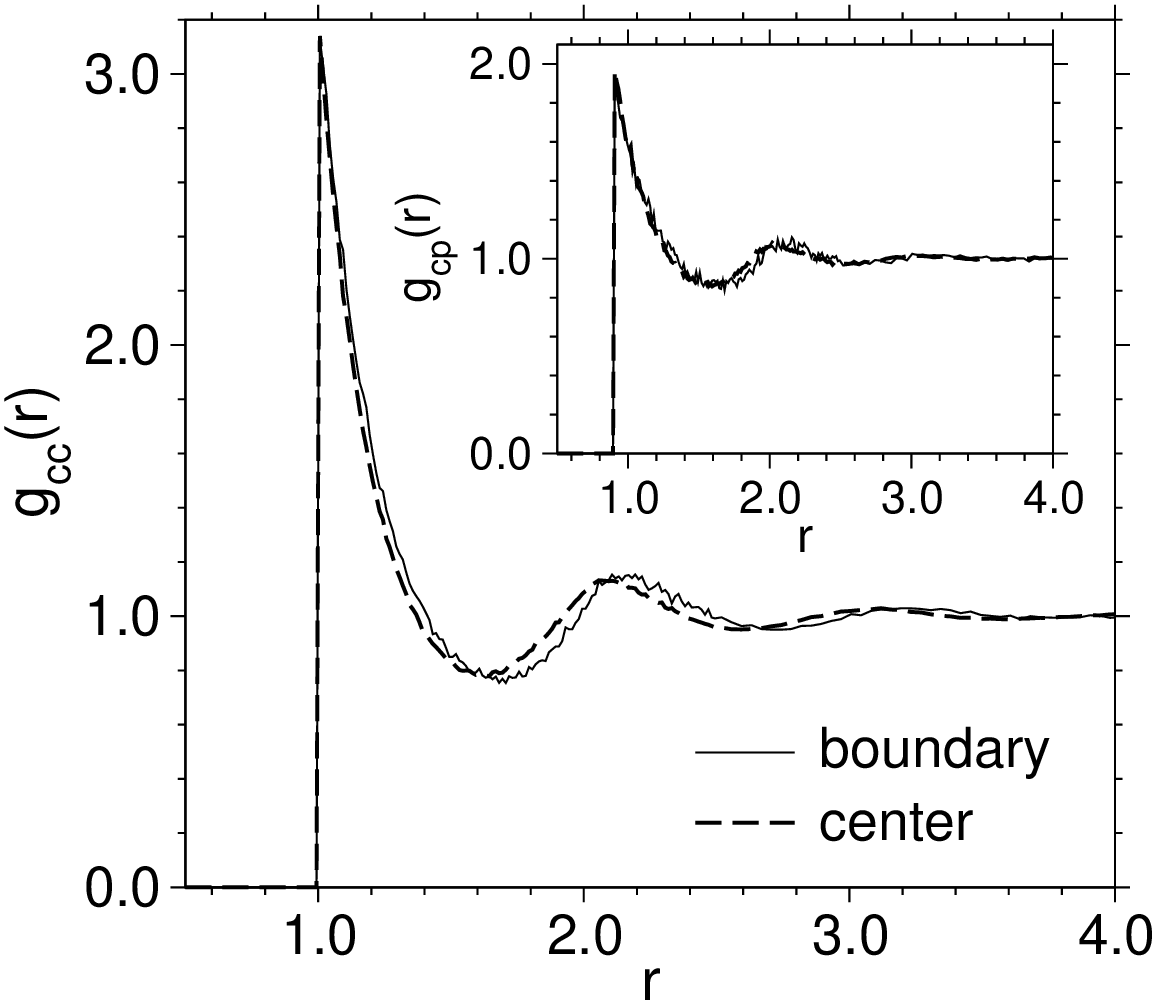}
\caption{\label{fig10} Radial distribution functions $g_{\rm cc}(r)$ (main 
frame) and $g_{\rm cp}(r)$ (inset) obtained in planes {\it parallel} to 
the confining walls (see details in text) for a film with thickness 
$D=10$. The full curves show data obtained in the direct vicinity of the 
walls; the broken curves show data obtained in the bulk region of the 
film. Here, the bulk region was taken to be in the range $3.5 \leq z \leq 
6.5$, where the corresponding density profile is essentially flat, see 
\fig{fig8}.}
\end{figure}

In view of the high colloid density reached locally in the first layer
adjacent to each wall, one may wonder whether this affects the lateral
ordering of the colloids in planes parallel to the walls. Therefore, we
have measured the radial distribution functions $g_{\alpha\beta}(r)$ for
colloid--colloid pairs ($\alpha\beta={\rm cc}$) and colloid-polymer pairs
($\alpha\beta={\rm cp}$), on the liquid branch of the binodal.  Here,
$r$ denotes the magnitude of the displacement vector $r=|\vec{r}_{ij}|$
between pairs of particles ($i,j={\rm c,p}$).  We have determined
$g_{\alpha\beta}(r)$ in the bulk region of the film (center), and in
the vicinity of the walls, i.e.~at a distance of $1\,\sigma$ from the
walls. Note that, in a system with walls (or virtual
boundaries that define, e.g., the center of the confined system),
the definition of the radial distibution function is different from that
in the bulk. In an isotropic system with periodic boundary conditions,
$g_{\alpha\beta}(r)$ is normalized by a phase factor $4\pi r^2$. However,
for the case that boundaries are present, this factor has to be modified
such that it is then determined only by that part of the surface $4\pi
r^2$ of a sphere around a particle that fits into the system. Thus,
the normalization factor depends on the distance of the particles from
the walls in $z$ direction.  The result for $g_{\alpha\beta}(r)$
is illustrated in \fig{fig10}, for thickness $D=10$, lateral dimension
$L=30$, and $\etapr=1.1$.  The important conclusion of \fig{fig10}
is that the radial distribution functions at the wall are almost
indistinguishable from those of the bulk! This shows that, at the value
of $\etapr$ considered here, there is still no sign of wall-induced
crystallization. Corresponding data for other film thicknesses, $D=3$
and $D=5$, have also been collected, but show the same trend as in
\fig{fig10}, and are therefore not shown.

\section{Influence of the colloid-wall interaction}

\begin{figure}
\centering
\includegraphics*[width=.4\textwidth]{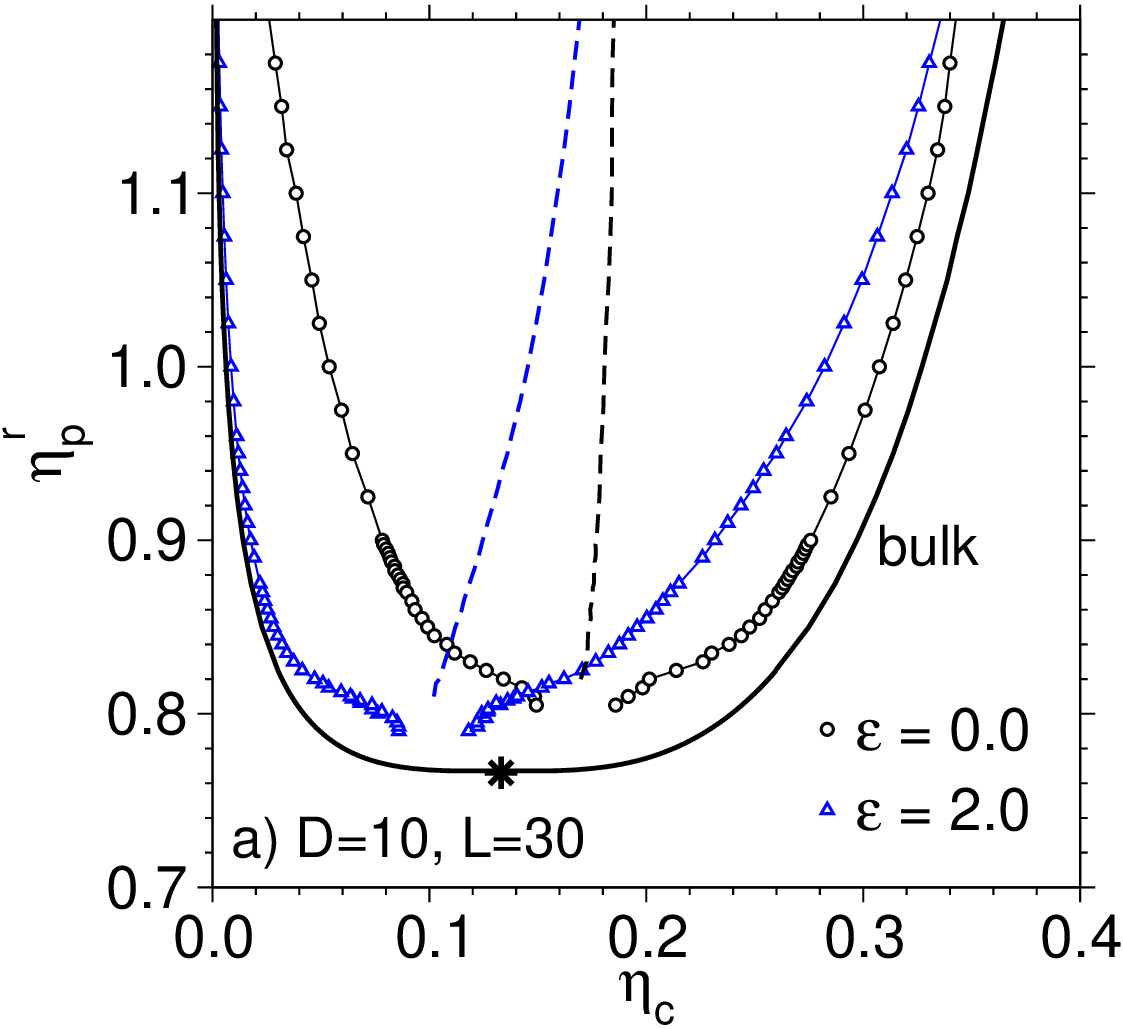}
\includegraphics*[width=.4\textwidth]{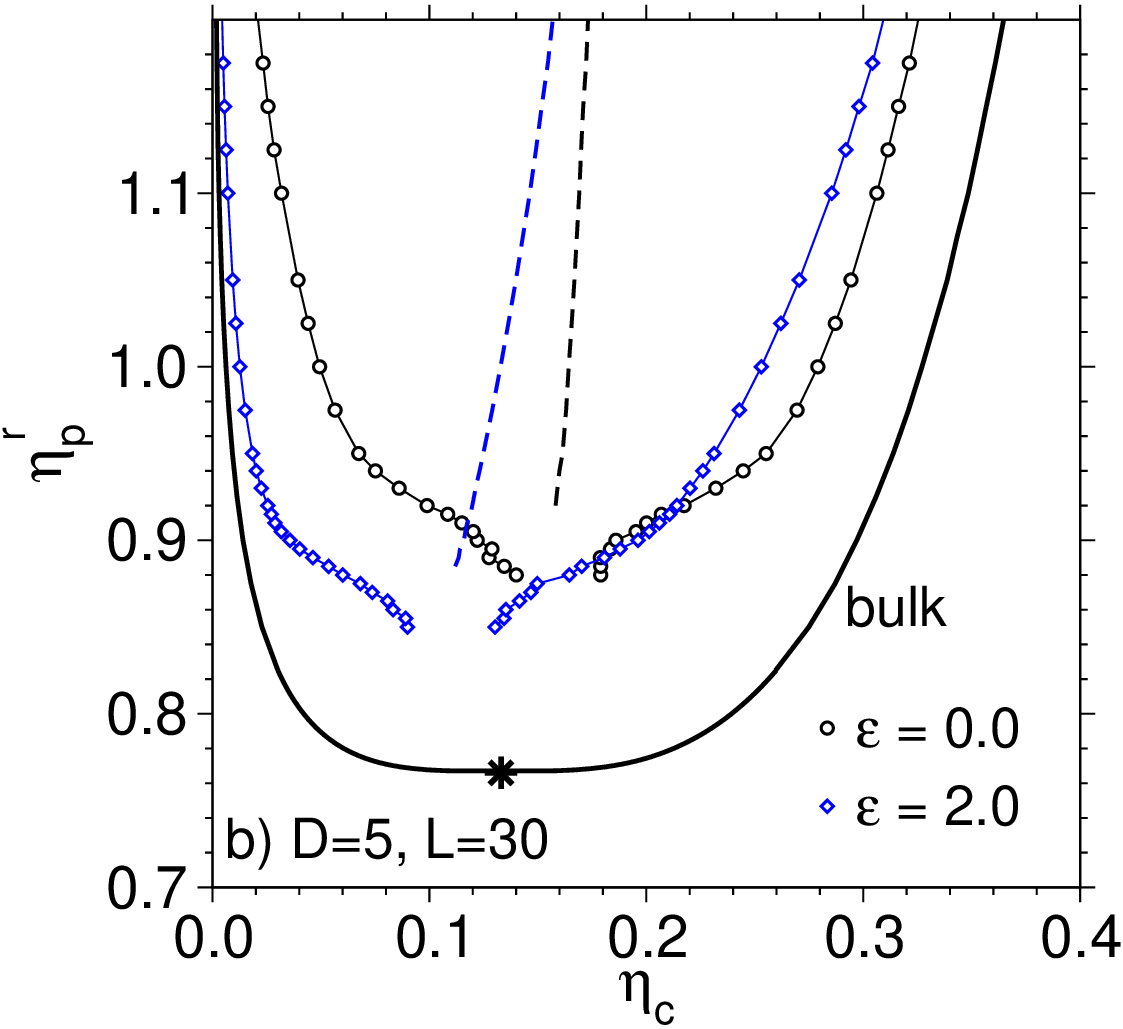}
\caption{\label{fig11} Binodal of the AO-model with $q=0.8$ in bulk (full 
curve) and confinement (symbols). Shown are data for two system sizes: 
$D=10,L=30$ (a) and $D=5,L=30$ (b). For each system size, two values of 
the colloid-wall interaction parameter $\epsilon$, see \eq{eq:cw}, are 
considered. Broken lines indicate the coexistence diameters of the 
confined systems.}
\end{figure}

\begin{figure}
\centering
\includegraphics*[width=.22\textwidth]{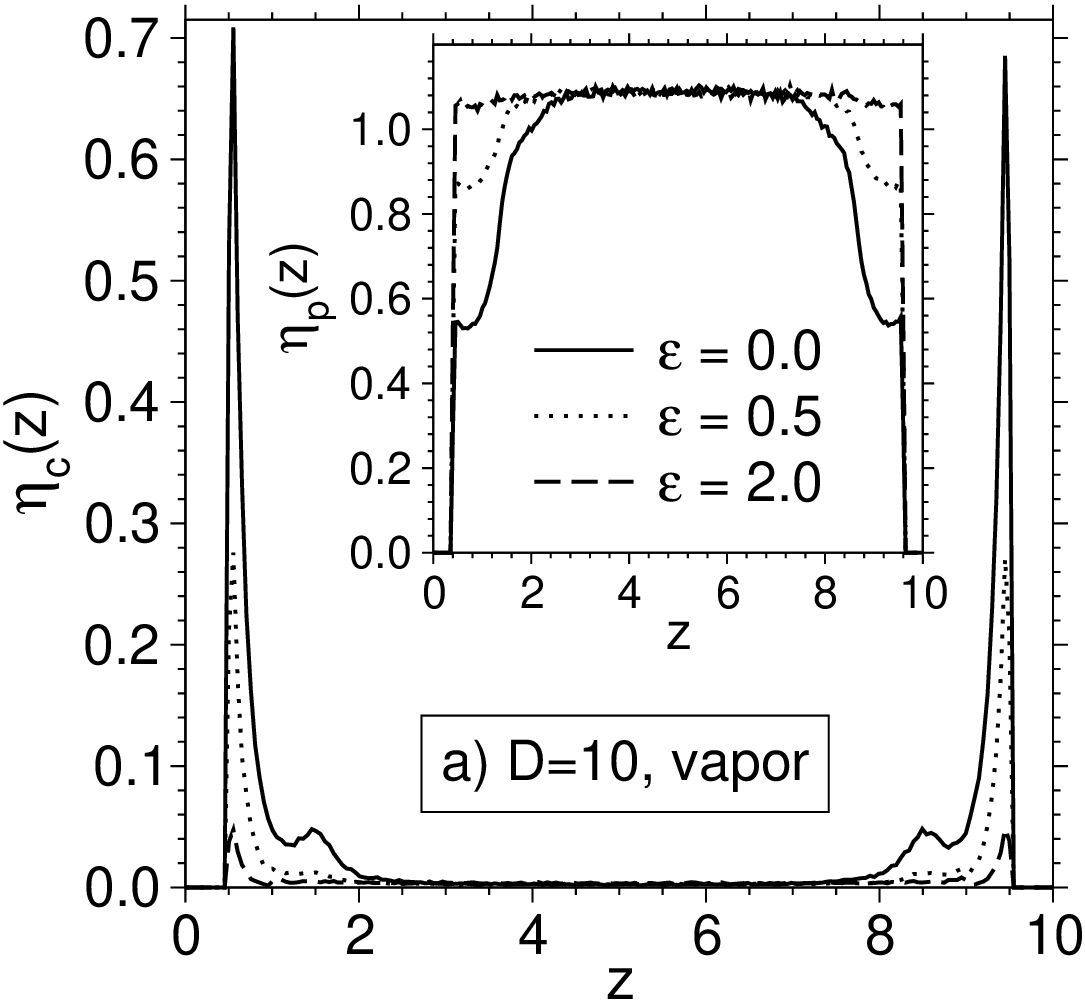}
\includegraphics*[width=.22\textwidth]{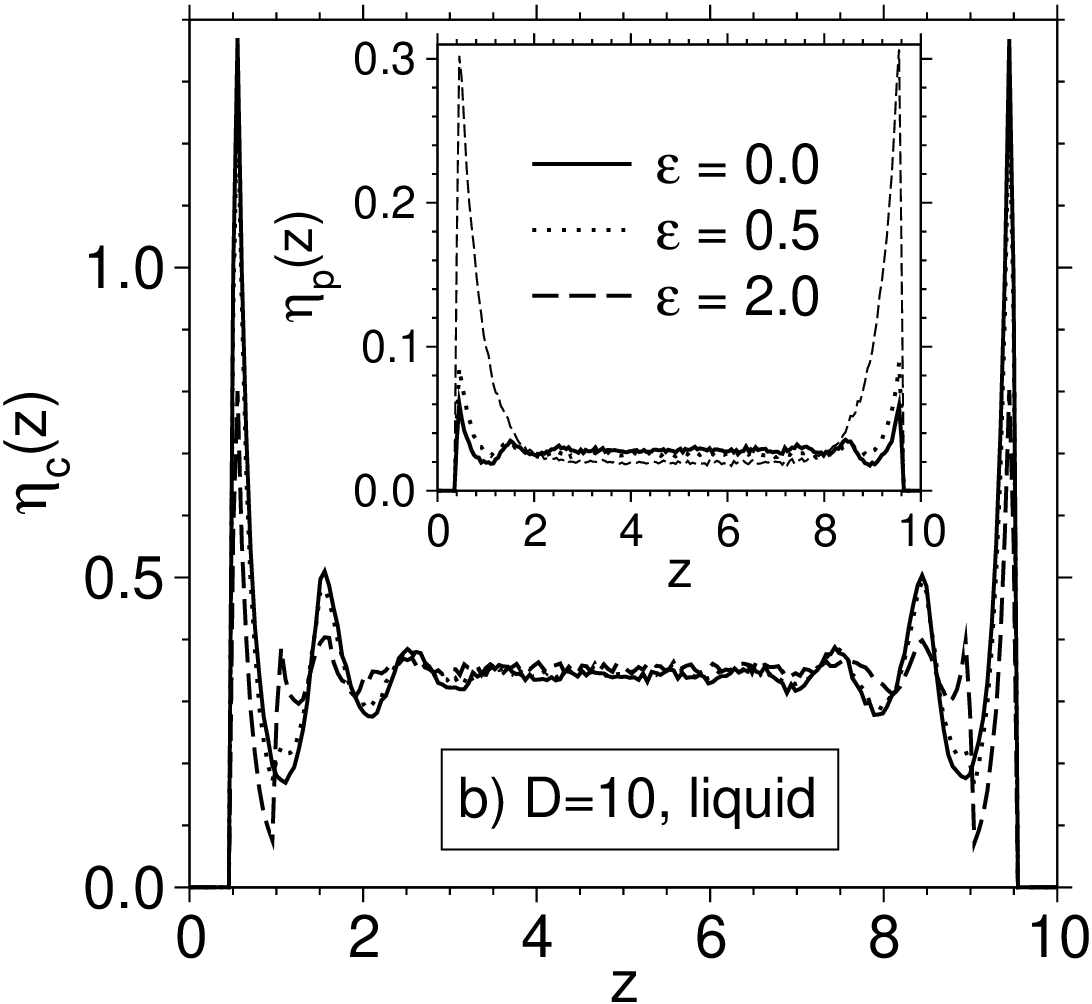}
\includegraphics*[width=.22\textwidth]{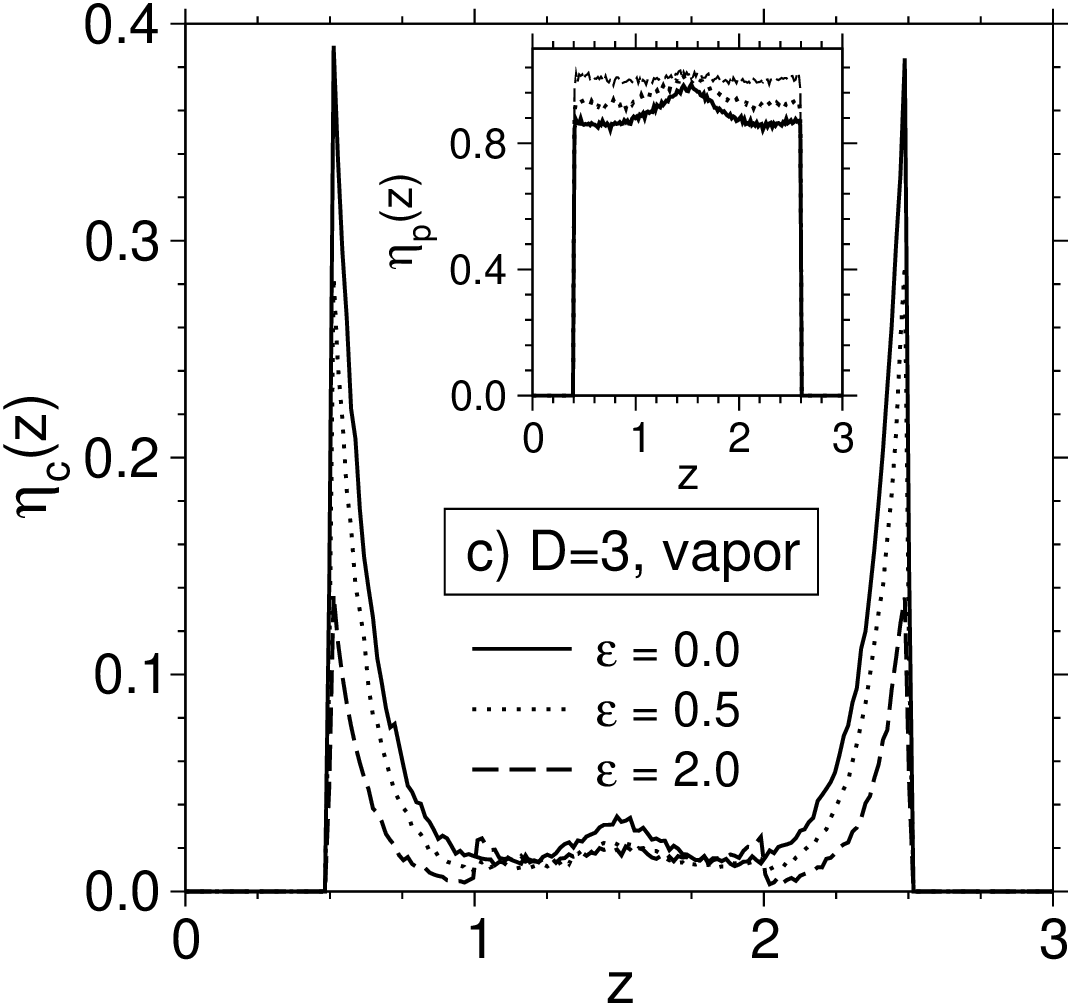}
\includegraphics*[width=.22\textwidth]{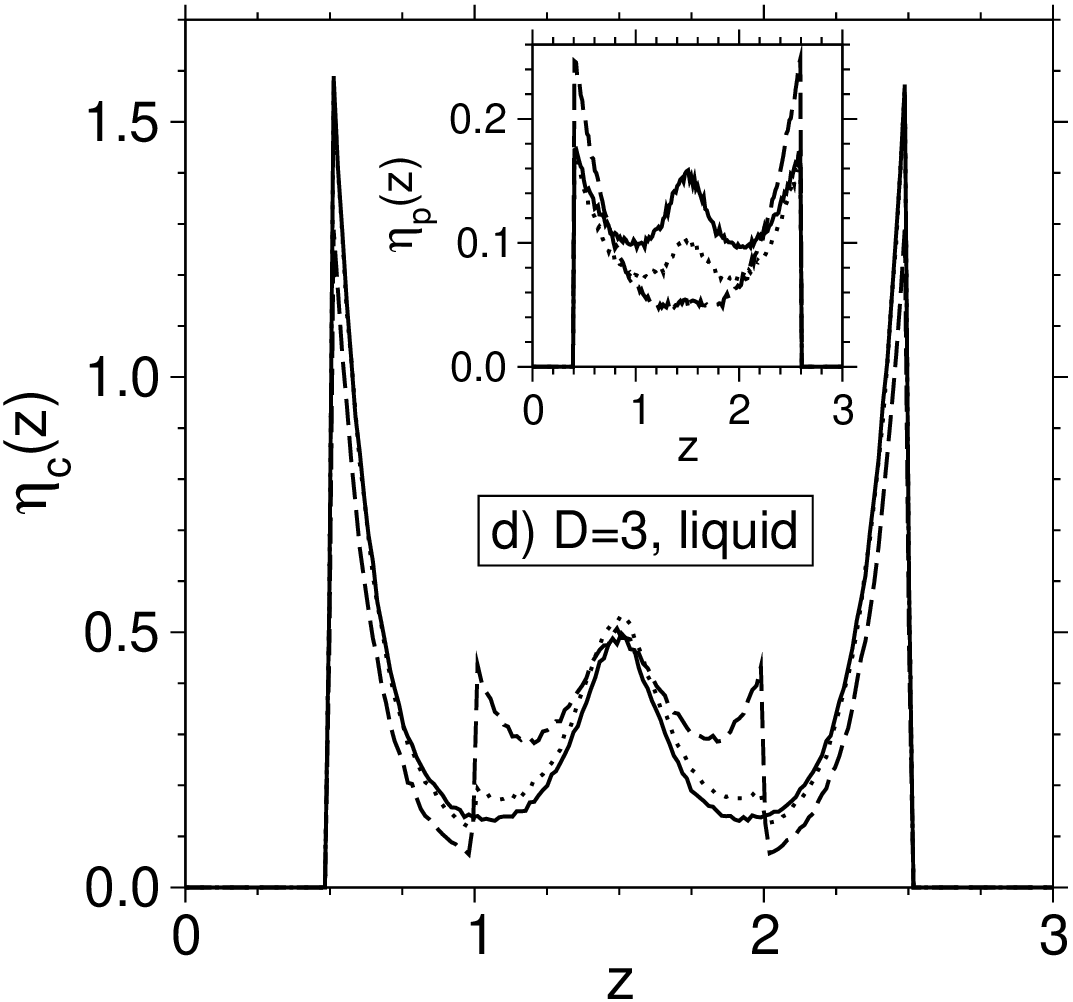}
\caption{\label{fig12} Colloid density profiles obtained in thin films at 
$\etapr=1.1$, for two values of the film thickness $D$, and several values 
of the colloid-wall parameter $\epsilon$ as indicated. Frames (a) and (b) 
show profiles obtained for $D=10$, on the vapor and liquid branch of the 
binodal, respectively. Frames (c) and (d) show the corresponding profiles 
for thickness $D=3$. The insets represent density profiles of the 
polymers.}
\end{figure}

\begin{figure}
\centering
\includegraphics*[width=.4\textwidth]{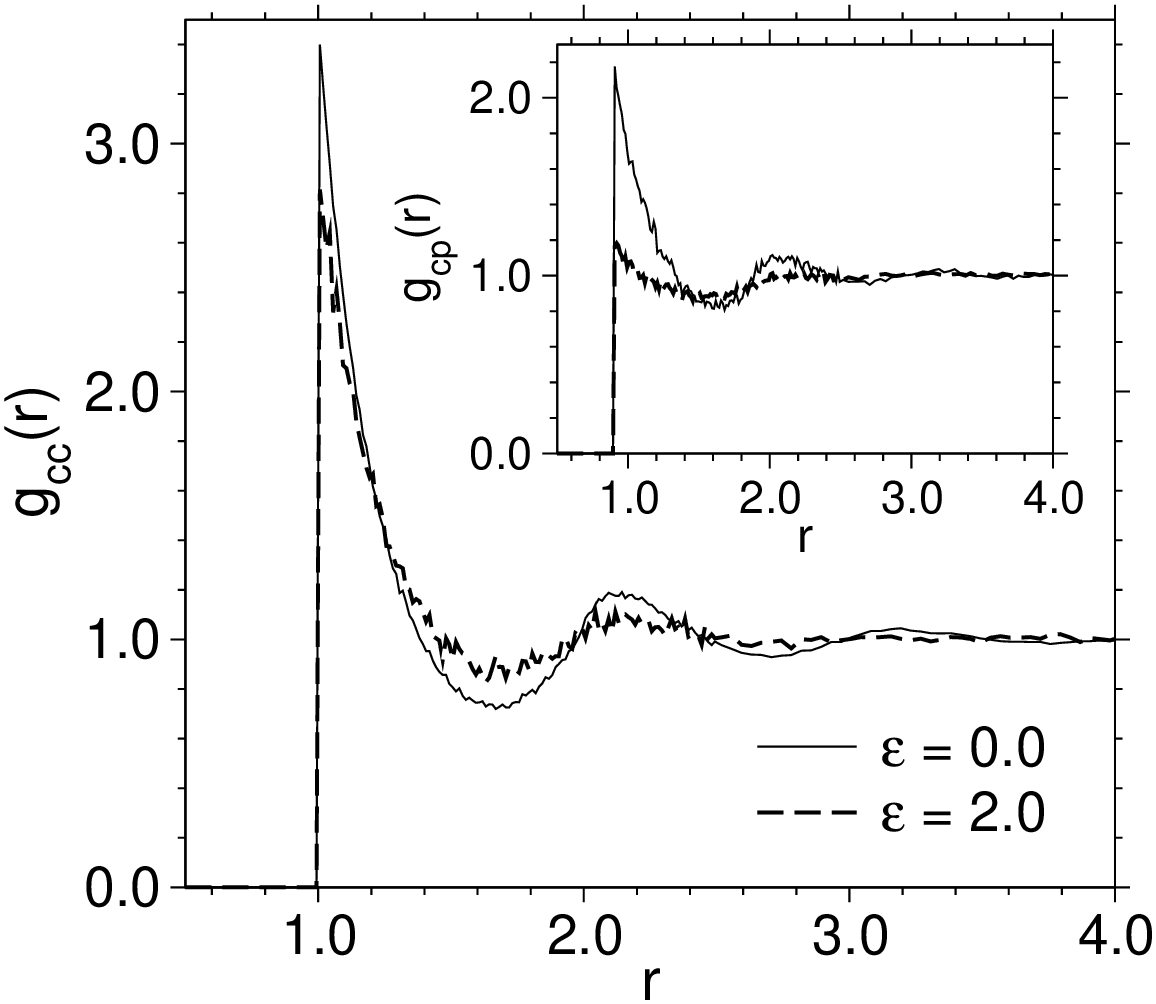}
\caption{\label{fig13} Radial distribution functions $g_{\rm cc}(r)$ (main 
frame) and $g_{\rm cp}(r)$ (inset), obtained in the boundary layer 
adjacent directly to the walls, for two values of $\epsilon$ as indicated. 
Results were obtained for thickness $D=10$, $L=30$, and $\etapr=1.1$.}
\end{figure}

The results so far were obtained using $\epsilon=0.5$ in the colloid-wall 
interaction of \eq{eq:cw}. In this section, the parameter $\epsilon$ 
itself is varied, and the corresponding changes in the phase behavior will 
be discussed. \fig{fig11} shows the binodal for thickness $D=10$ and 
$D=5$, for two different values of $\epsilon$, compared to the bulk 
binodal. As before, $\epsilon=0$ implies pure hard walls \cite{8}. We 
observe that, for large $\etapr$, the binodal corresponding to 
$\epsilon=2.0$ and $D=10$, approaches the bulk binodal rather closely. 
This indicates that $\epsilon=2.0$ approximately cancels the entropic 
colloid-wall attraction. The latter is also evident from the corresponding 
density profiles, see Figs.~\ref{fig12}(a) and~(b). For $\epsilon=2.0$, 
the peak in the colloid density profile in the layers adjacent to the 
walls, is significantly reduced. In the liquid phase, we instead recognize 
a rather strong density jump for $\epsilon=2.0$, at both $z=1$ and 
$z=D-1$, due to the step in the colloid-wall potential. Only for 
$\epsilon=0$, do the density profiles remain smooth.

Clearly, in future work, a careful analysis of more realistic colloid-wall 
interactions is desirable. Such a study is moreover warranted because we 
find, see \fig{fig13}, that variations in $\epsilon$ significantly 
influence the radial distribution functions, $g_{\rm cc}(r)$ and $g_{\rm 
cp}(r)$, measured at the walls. By increasing $\epsilon$, a reduction of 
the first peak in $g_{\rm cc}(r)$ occurs, with the second peak becoming 
almost completely washed out; for $g_{\rm cp}(r)$ the effect is even more 
dramatic. As expected, increasing $\epsilon$ reduces the colloid density 
at the walls somewhat, thereby also weakening their tendency to order in 
the layers adjacent to the walls. Similar conclusions emerge from other 
wall thicknesses, too, and are therefore not shown here.

\section{Discussion and concluding remarks}

Colloid--polymer mixtures are important model systems for the experimental 
exploration of fluid--fluid phase separation \cite{19,20} and interfacial 
phenomena \cite{21,22}. In order to guide possible future experiments on 
phase separation of such systems in confined geometry, we have carried out 
a simulation study of a generic model system, the AO model, confined to a 
slit formed by two parallel walls. The walls were not only impenetrable to 
both colloids and polymers, but in addition a short--ranged repulsive 
interaction between the walls and the colloids was added. Such an 
interaction can be realized, for instance, by a suitable coating of the 
walls with a moderately dense polymer brush (with the same chemical 
composition of the polymers dissolved in the solution). In this way, the 
strength of the depletion interaction between the colloid and the walls 
can be modified.

Our results show that the unmixing binodal, describing phase separation in 
colloid--polymer mixtures, changes dramatically from its bulk form, when 
the width of the pore is in the range of 3--10 colloid diameters. While 
such ultra--thin pores seem to be experimentally realizable for colloidal 
systems, where the particle diameters are in the micrometer range, a 
corresponding study of a mixture of small molecules or atoms, clearly, 
would be extremely difficult, if not impossible. Surprisingly, even for 
thin pores, we find that bulk--like phase behavior in the center of the 
pore is obtained, provided one is far away from criticality. Only for 
pores with thicknesses smaller than $D=3\sigma_{\rm c}$ or so, can a 
bulk--like region in the center of the pore no longer be identified. In 
this case, the layering induced in the density by the left wall interferes 
with the layering due to the right wall. Therefore, such extremely thin 
slit pores cannot be accounted for quantitatively in terms of a 
decomposition of their properties into bulk and surface properties. Hence, 
the description of the shift of the transition between the phases poor in 
colloids (``vapor'') and rich in colloids (``liquid'') in terms of the 
Kelvin equation, no longer is accurate. Nevertheless, agreement with the 
Kelvin equation is already recovered again above $D=10\sigma_{\rm c}$ or 
so, which is still remarkably thin.

The main emphasis of this work, however, has been a study of the shift of 
the critical point. It has been shown that the scaling theory of Fisher 
and Nakanishi \cite{5}, originally formulated for the Ising lattice gas 
model exhibiting full particle--hole symmetry between the coexisting 
phases, is compatible with our present numerical results. A nontrivial 
finding, which needs further investigation, is the fact that the critical 
value of the coexistence diameter approaches its limiting value in a 
non-monotonic way when $D\to\infty$.

In our study, we have obtained not only accurate binodal curves of 
confined colloid--polymer mixtures, but we have also characterized the 
coexisting colloid--rich and colloid--poor phases in terms of their 
concentration profiles across the slit pore, and in terms of their radial 
distribution functions obtained in planes parallel to the walls. It would 
be interesting to extract corresponding information from experiments. 
Since the AO model is a crude simplification of reality, characteristic 
differences between simulation and experiment are to be expected, 
elucidating the limitations of the AO model.

In future work, we plan to extend our study to confinement between two 
parallel planar walls which are nonequivalent, whereby, for example, only 
one wall is coated by the polymer brush. In this way, an experimental 
realization of the interface localization--delocalization transition 
\cite{4} may become possible. Also an extension of such studies to the 
dynamics of phase separation under confinement is of interest. We hope 
that the present work will stimulate corresponding experimental efforts.

{\bf Acknowledgments:} This work received financial support from the 
Deutsche Forschungsgemeinschaft (DFG) through the SFB-TR6, subprojects A5 
and D3. Stimulating discussions with M. Dijkstra and H. N. W. Lekkerkerker 
are also acknowledged.


\begin{thebibliography}{99}
%
\bibitem{1} 
D. E. Sullivan and M. M. Telo da Gama, 
in \textit{Fluid Interfacial Phenomena}, edited by 
C. A. Croxton, p. 45
(Wiley, New York, 1985).
%
\bibitem{2} 
S. Dietrich, 
in \textit{Phase Transitions and Critical Phenomena, Vol.~12},
edited by C. Domb and J. L. Lebowitz, p. 1
(Academic Press, New York, 1988).
%
\bibitem{3} 
M. Schick, 
in \textit{Liquids at Interfaces}, 
edited by J. Charvolin, J.-F. Joanny, and J. Zinn-Justin, p. 415
(North-Holland, Amsterdam, 1990).
%
\bibitem{4} 
K. Binder, D. P. Landau, and M. M\"uller, 
J. Stat. Phys. \textbf{110}, 1411 (2003).
%
\bibitem{5} 
M. E. Fisher and H. Nakanishi, 
J. Chem. Phys. \textbf{75}, 5857 (1981); 
H. Nakanishi and M. E. Fisher, 
J. Chem. Phys. \textbf{78}, 3279 (1983).
%
\bibitem{6} 
R. Evans, 
J. Phys.:  Condens. Matter \textbf{2}, 8989 (1990).
%
\bibitem{7} 
L. D. Gelb, K. E. Gubbins, R. Radhakrishnan, and M. Sliwinska-Bartkowiak, 
Rep. Prog. Phys. \textbf{62}, 1573 (1999).
%
\bibitem{8} 
R. L. C. Vink, K. Binder, and J. Horbach, 
Phys. Rev. E {\bf 73}, 056118 (2006).
%
\bibitem{9} 
A more complete bibliography on the extensive literature on 
the subject can be found in Refs. 7, 8.
%
\bibitem{10} 
Y. Rouault, J. Baschnagel, and K. Binder, 
J. Stat. Phys. \textbf{80}, 1009 (1995).
%
\bibitem{11} 
O. Dillmann, W. Janke, M. M\"uller and K. Binder, 
J. Chem. Phys. \textbf{114}, 5853 (2001).
%
\bibitem{12} 
F. Rouquerol, J. Rouquerol, and K. S. W. Sing,
\textit{Adsorption by Powders and Porous Solids: Principles,
Methodology and Applications} 
(Academic Press, San Diego, 1999).
%
\bibitem{13} 
R. Kelsall, I. W. Hamley and M. Geoghegan (eds.),
\textit{Nanoscale Science and Technology} 
(Wiley-VCH, Weinheim, 2005).
%
\bibitem{14}
T. M. Squires and S. R. Quake,
Rev. Mod. Phys. {\bf 77}, 977 (2005).
%
\bibitem{15} 
A. Patrykiejew, S. Sokolowski, and K. Binder, 
Surf. Sci. Rep. \textbf{37}, 207 (2000).
%
\bibitem{16} 
W. C. Poon and P. N. Pusey, 
in \textit{Observation, Prediction and Simulation of Phase 
Transitions in Complex Fluids}, p. 3, edited by M. Baus, L. F. Rull, 
and J. P. Ryckaert (Kluwer Acad. Publ., Dordrecht, 1995).
%
\bibitem{17} 
A. K. Arora and B. V. R. Tata, 
Adv. Colloid Interface Sci. \textbf{78}, 49 (1998).
%
\bibitem{18} 
H. L\"owen, 
J. Phys.: Condens. Matter \textbf{13}, R415 (2001).
%
\bibitem{19} 
W. C. Poon, 
J. Phys.: Condens. Matter \textbf{14}, R589 (2002).
%
\bibitem{20} 
D. G. A. L. Aarts, J. H. van der Wiel, and H. N. W. Lekkerkerker, 
J. Phys.: Condens. Matter \textbf{15}, S245 (2003).
%
\bibitem{21} 
D. G. A. L. Aarts, M. Schmidt, and H. N. W. Lekkerkerker, 
Science \textbf{304}, 847 (2004).
%
\bibitem{22} 
D. G. A. L. Aarts and H. N. W. Lekkerkerker, 
J. Phys.: Condens. Matter \textbf{16}, S4231 (2004).
%
\bibitem{23} 
S. Asakura and F. Oosawa, 
J. Chem. Phys. \textbf{22}, 1255 (1954).
%
\bibitem{24} 
A. Vrij, 
Pure Appl. Chem. \textbf{48}, 471 (1976).
%
\bibitem{25} 
M. Dijkstra and R. van Roij, 
Phys. Rev. Lett. \textbf{89}, 208303 (2002).
%
\bibitem{26} 
M. Dijkstra and R. van Roij, 
J. Phys.: Condens. Matter \textbf{17}, S3507 (2005).
%
\bibitem{27} 
M. Schmidt, A. Fortini, and M. Dijkstra, 
J. Phys.: Condens. Matter \textbf{15}, S3411 (2003); 
M. Schmidt, A. Fortini, and M. Dijkstra, 
J. Phys.: Condens. Matter \textbf{16}, S4159 (2004).
%
\bibitem{28} 
A. Fortini, M. Schmidt, and M. Dijkstra,
Phys. Rev. E {\bf 73}, 051502 (2006).
%
\bibitem{29} 
R. L. C. Vink and J. Horbach, 
J. Chem. Phys. \textbf{121}, 3253 (2004); 
R. L. C. Vink, 
in \textit{Computer Simulation Studies in Condensed Matter Physics XVII}, 
edited by D. P. Landau, S. P. Lewis, and H. B. Sch\"uttler 
(Springer, Berlin, 2004).
%
\bibitem{30} 
R. L. C. Vink and J. Horbach, 
J. Phys.: Condens. Matter \textbf{16}, S3807 (2004); 
R. L. C. Vink, J. Horbach, and K. Binder, 
J. Chem. Phys. \textbf{122}, 134905 (2005).
%
\bibitem{31} 
M. E. Fisher, 
Rev. Mod. Phys. \textbf{46}, 587 (1974).
%
\bibitem{32} 
M. E. Fisher and S.-Y. Zinn, 
J. Phys. A: Math. Gen. \textbf{26}, 201 (1998).
%
\bibitem{33} 
J. Zinn-Justin, 
Phys. Rep. \textbf{344}, 159 (2001).
%
\bibitem{34} 
K. Binder and E. Luijten, 
Phys. Rep. \textbf{344}, 179 (2001).
%
\bibitem{35} 
R. J. Baxter, 
\textit{Exactly Solved Models of Statistical Mechanics} 
(Academic Press, London, 1982).
%
\bibitem{36} 
W. T. Thomson (Lord Kelvin), 
Philos. Mag. \textbf{42}, 448 (1871).
%
\bibitem{37} 
K. Binder and P. C. Hohenberg, 
Phys. Rev. B \textbf{6}, 3461 (1972); 
ibid \textbf{9}, 2192 (1974).
%
\bibitem{38} 
K. Binder, 
in \textit{Phase Transitions and Critical Phenomena, Vol. 8} 
(C. Domb and J. L. Lebowitz, eds.)
(Academic Press, London 1983) p. 1.
%
\bibitem{39} 
K. Binder and D. P. Landau, 
Phys. Rev. Lett. \textbf{52}, 318 (1984); 
D. P. Landau and K. Binder, 
Phys. Rev. B \textbf{41}, 4633 (1990).
%
\bibitem{40} 
C. Ruge, S. Dunkelmann, F. Wagner, and J. Wulf, 
J. Stat. Phys. \textbf{73}, 293 (1993); 
C. Ruge and F. Wagner, 
Phys. Rev. B \textbf{52}, 4209 (1995).
%
\bibitem{41} 
H. W. Diehl and M. Shpot, 
Nucl. Phys. B \textbf{528}, 595 (1998).
%
\bibitem{42} 
N. B. Wilding, 
in \textit{Annual Reviews of Computational Physics}, 
edited by D. Stauffer (World Scientific,
Singapore, 1996) p. 37.
%
\bibitem{43} 
I. Brovchenko, A. Geiger, and A. Oleinikova, 
Phys. Chem. Chem. Phys. \textbf{3}, 1567 (2001); 
J. Phys.: Condens. Matter \textbf{16}, S5345 (2004).
%
\bibitem{44} 
I. Brovchenko, A. Geiger, and A. Oleinikova, 
Eur. Phys. J. B \textbf{44}, 345 (2005); 
A. Oleinikova, I. Brovchenko, and A. Geiger, 
J. Phys.: Condens. Matter \textbf{17}, 7845 (2005).
%
\bibitem{45} 
A. Oleinikova, I. Brovchenko, and A. Geiger, 
Eur. Phys. J., in press.
%
\bibitem{46} 
P. G. de Gennes, 
\textit{Scaling Concepts in Polymer Physics} 
(Cornell Univ. Press, Ithaca, N. Y., 1979).
%
\bibitem{47} 
P. Virnau and M. M\"uller, 
J. Chem. Phys. \textbf{120}, 10925 (2004).
%
\bibitem{48} 
K. Binder and D. P. Landau, 
Phys. Rev. B \textbf{30}, 1477 (1984).
%
\bibitem{49} 
C. Borgs and S. Kappler, 
Phys. Lett. A \textbf{171}, 37 (1992); 
C. Borgs and R. Kotecky, 
J. Stat. Phys. \textbf{61}, 79 (1990).
%
\bibitem{50} 
K. Binder and D. W. Heermann, 
\textit{Monte Carlo Simulation in Statistical Physics: An 
Introduction, 4th Edition} (Springer, Berlin, 2002).
%
\bibitem{51} 
D. P. Landau and K. Binder, 
\textit{A Guide to Monte Carlo Simulation in Statistical 
Physics, 2nd ed.} (Cambridge Univ. Press, Cambridge, 2005).
%
\bibitem{52} 
K. Binder, 
Z. Phys. B \textbf{43}, 119 (1981).
%
\bibitem{53} 
Y. C. Kim, M. E. Fisher and E. Luijten, 
Phys. Rev. Lett. \textbf{91}, 65701 (2003); 
Y. C. Kim and M. E. Fisher,
Comp. Phys. Comm. \textbf{169}, 295 (2005); 
Y. C. Kim, 
Phys. Rev. E \textbf{71}, 051501 (2005).
%
\bibitem{54} 
G. Kameniarz and H. W. J. Bl\"ote, 
J. Phys. A: Math. Gen. \textbf{26}, 201 (1993).
%
\bibitem{55}
Y. C. Kim, M. E. Fisher, and G. Orkoulas,
Phys. Rev. E {\bf 67}, 061506 (2003).
%
\bibitem{56}
A. Anisimov, private communication; R. L. C. Vink and H. H. Wensink, 
submitted (2006).
%
\end{thebibliography}
\end{document}